\documentclass[11pt]{article}

\usepackage[margin=1in]{geometry}
\usepackage[T1]{fontenc}
\usepackage[utf8]{inputenc}
\usepackage{lmodern}

\usepackage{graphicx}
\usepackage{caption}
\usepackage{subcaption}
\usepackage{booktabs}
\usepackage{amsmath,amssymb}
\usepackage{textcomp}

\usepackage{appendix}
\usepackage{ragged2e}
\usepackage{authblk}

\usepackage{natbib}

\usepackage[colorlinks=true,citecolor=blue,linkcolor=blue,urlcolor=blue]{hyperref}

\graphicspath{{./}{Figure/}}

\newcommand{\arcsec}{\ensuremath{^{\prime\prime}}}
\newcommand{\arcmin}{\ensuremath{^{\prime}}}
\newcommand{\sun}{\ensuremath{\odot}}

\newlength{\tabnotewidth}
\newcommand{\tablecols}[1]{}%
\newcommand{\tabnotemark}[1]{\textsuperscript{#1}}%
\newcommand{\tabnotetext}[2]{\par\noindent\tabnotemark{#1}~#2\par}%

\newcommand{\keywords}[1]{\par\medskip\noindent\textbf{Keywords:} #1\par\medskip}

\newenvironment{resumen}{%
  \begingroup
  \begin{abstract}%
}{%
  \end{abstract}%
  \endgroup
}

\title{Physical characterization and modeling of candidate Hyper-Compact HII Regions}

\date{}

\author[1,2]{I.~T.~Rodr\'\i{}guez-Esnard\thanks{Corresponding author: it.rodriguez@ugto.mx}}
\author[3]{S.~Kurtz}
\author[4]{J.~D.~Pandian}
\author[5]{J.~Franco}
\author[6,7]{A.~S\'anchez-Monge}
\author[1]{M.~A.~Trinidad}
\author[8]{V.~Migenes}

\affil[1]{Departamento de Astronom\'\i{}a, Universidad de Guanajuato, Guanajuato, M\'exico.}
\affil[2]{Instituto de Cibern\'etica, Matem\'atica y F\'\i{}sica, Cuba}
\affil[3]{Instituto de Radioastronom\'\i{}a y Astrof\'\i{}sica, Universidad Nacional Aut\'onoma de M\'exico, M\'exico.}
\affil[4]{Department of Earth \& Space Science, Institute of Space Science \& Technology, India.}
\affil[5]{Instituto de Astronom\'\i{}a, Universidad Nacional Aut\'onoma de M\'exico, M\'exico.}
\affil[6]{Institute of Space Sciences (ICE-CSIC), Barcelona, Spain}
\affil[7]{Institute of Space Studies of Catalonia (IEEC), Barcelona, Spain}
\affil[8]{Department of Physics, Texas Southern University, Texas, USA}

\begin{document}

\maketitle
\thispagestyle{plain}

\begin{abstract}
Hypercompact HII regions (HC) are regions of ionized gas associated with the early stages of high-mass star formation. With the aim of better understanding their characteristics, we studied five candidate HC HII regions. Here, we present observations with the Jansky Very Large Array (VLA) at 2 and 6 cm, with angular resolutions in the range of $\sim$1 -- 3\arcsec and report the images of the detected sources and the measured parameters. In addition, we explore several possible scenarios, considering the regions as both uniform and non-uniform spheres, and as winds, both spherical and collimated. In most cases, the sources were unresolved, but by applying the models, we estimate that their sizes vary in a range of 0.3 to 3.7 mpc while their electron densities are in the range of $1.3 \times 10^{5}$ to $2.4 \times 10^{6}$ cm$^{-3}$, indicating that most sources are consistent with small, weak UC HII regions, although a few remain viable candidates for HC HII regions, with G40.28$-$0.22 as the strongest case. We do not rule out the possibility that some sources are jets or stellar winds.
\end{abstract}

\keywords{HII regions, Stars: formation, Stars: massive}

\begin{resumen}
Las regiones HII hipercompactas son regiones de gas ionizado asociadas con las etapas tempranas de formaci\'on de estrellas de alta masa. Realizamos un estudio de cinco candidatos a regiones HII HC, con el objetivo de comprender mejor sus caracter\'\i sticas. Para esto se llevaron a cabo observaciones con el Jansky Very Large Array (VLA) en 2 y 6 cm, con resoluciones angulares en el rango de $\sim$1 -- 3 \arcsec y reportamos las im\'agenes de las fuentes detectadas y los par\'ametros medidos. Adem\'as, exploramos varios escenarios posibles, considerando las regiones como esferas (uniformes y no-uniformes) y como vientos (esf\'ericos tanto como colimados). En general, las fuentes fueron no-resueltas, sin embargo, al aplicar los modelos, hemos podido estimar que sus tama\~nos var\'\i an en un rango de 0.3 a 3.7 mpc y que las densidades electr\'onicas est\'an en el rango de $1.3 \times 10^{5}$ a $2.4 \times 10^{6}$ cm$^{-3}$, lo que indica que la mayor\'\i a de las fuentes son consistentes con peque\~nas y d\'ebiles regiones UC HII, aunque algunas siguen siendo candidatas viables a regiones HC HII, destacando G40.28$-$0.22 como el caso m\'as fuerte. No descartamos la posibilidad de que algunas fuentes sean chorros o vientos estelares.
\end{resumen}


\section{Introduction}
\label{sec.1}

Hypercompact (HC) HII regions are a class of ionized gaseous nebulae, linked to the early phases of high-mass star formation. They are commonly located within dusty clumps, associated with  masers, outflows, broad radio recombination lines, and extended green objects \citep{Yang2021}. These regions have  distinctive properties that set them apart from other HII regions, such as compact and ultracompact HII regions: they are smaller ($\lesssim$0.05 pc), with higher electron densities (n$_{e}\gtrsim$10$^{6}$ cm$^{-3}$) and extremely high emission measures (EM$\gtrsim$10$^{9}$ pc cm$^{-6}$) \citep{Kurtz2002}.

It is expected that HC HII regions are optically thick at centimeter wavelengths. This characteristic leads to a positive spectral index ($\alpha$, \citet{Kurtz2005}) indicating that the observed brightness increases with frequency.  There are different theories to account for the properties of HC HII regions with an intermediate spectral index ($\alpha \sim$ +1) in the optically thick regime rather than the canonical value of $+2$ for an optically thick, uniform density region. It has been proposed that they are composed of multiple unresolved clumps of gas \citep{Ignace2004} or that they have density gradients \citep{Franco2000}.

The small sizes of HC HII regions can be explained by models of photo-evaporating disk winds \citep{Hollenbach1994, Lizano1996, Lugo2004}. Alternatively, HC HII regions can be explained by ionized accretion flows, where the stellar mass gradually increases due to accretion through the HC HII region \citep{Keto2003}. \citet{Tan2003} proposed that high accretion rates for massive young stellar objects lead to higher outflow rates, creating small, jet-like HC HII regions confined by their outflows. The extremely broad radio recombination lines observed in some HC HII regions can be attributed to bulk gas motions such as accretion, rotation, or expansion, with some contribution from pressure broadening \citep{Lizano2008}.

Unfortunately, there is no large sample of known hypercompact (HC) HII regions to study their properties in a statistically significant way.  Some progress in this regard has recently been made by two groups.  \citet{Yang2021} identified 16 HC HII regions in a sample of 120 HII regions, while \citet{Patel2023,Patel2024,Patel2025} report 46 confirmed and/or candidate HC HII regions. Our previous studies discovered six candidate HC HII regions within a sample of 24 HII regions using an angular resolution of 6\arcsec and 9\arcsec  \citep{Sanchez2011}. However, they were not able to spatially resolve the sources and place strong constraints on their properties. They inferred that these regions are very compact, with sizes between 1 and 10 mpc and spectral indices between 0.4 and 1.4. In this paper, we investigate five HC HII region candidates from \citet{Sanchez2011} by conducting new observations at 2 and 6~cm to better constrain the source properties.

We outline the observations and the data reduction methods in section $\S$\ref{sec.2}. Section $\S$\ref{sec.3} reports our observational results. Section $\S$\ref{sec.4} applies various models to estimate the physical parameters of the sources, and section $\S$\ref{sec.5} discusses our findings for each source. In section $\S$\ref{sec.6} we discuss our main results and summarize the main conclusions of our study.

\section{Observations}
\label{sec.2}

We observed five candidate HC HII regions, which were selected from \citet{Sanchez2011} because they show 6.7 GHz methanol maser emission (a tracer of high-mass star formation) and they show little or no emission at 6 or 20 cm (suggesting a pre-ultracompact HII region stage). All sources show extended 8 and 24 $\mu$m emission, further supporting their identification as very young high-mass star-forming regions. The sources are listed in Table \ref{table0}.

The observations were made with the Jansky Very Large Array (VLA) of the NRAO\footnote{The National Radio Astronomy Observatory (NRAO) is a facility of the National Science Foundation operated under cooperative agreement by Associated Universities, Inc.} in the 6 and 2 cm (5.8 and 14.5 GHz) continuum emission using the B and C configurations, respectively, during two observing runs in February and June 2012 (project code VLA/12A-145). The observations were made in scaled-arrays, so the synthesized beams were expected to be approximately the same ($\sim$ 0.9\arcsec) at both wavelengths. However, during the 2 cm C-array observation, about one-third of the antennas did not have 2 cm receivers installed, resulting in a larger beam of $\sim$2\arcsec.

The observations at both wavelengths were made with a bandwidth of 2048 MHz comprised of  16 spectral windows, each containing 128 1-MHz channels at 6 cm and 64 2-MHz channels at 2 cm. Right and left circular polarizations were sampled,  and the on-source integration time for each field was 10 minutes.

The amplitude and bandpass calibrator used was J1331+305 (3C286) with a calculated flux density of 6.4 Jy at  6 cm and 3.6 Jy at 2 cm using the  Perley and Butler 2010 scale. The phase calibrators were J1851+0035 (for G34.82+0.35 and G40.28-0.22)  with flux densities of $\sim$0.8 Jy and $\sim$ 1.0 Jy at 6 cm and 2 cm, respectively and J1922+1530 (for G48.99-0.30, G53.04+0.11, and G53.14+0.07) with flux densities of $\sim$0.4 Jy and $\sim$0.3 Jy at 6 cm and 2 cm, respectively.

In addition, we used previous  observations from \citet{Sanchez2011}. To ensure a uniform treatment of all the data, we reduced their observations at 3.6 and 1.3 cm using the same procedures as for the new data at 2 and 6 cm. Calibration and imaging were performed using the software CASA. The observing parameters of the maps (beam size and rms noise), as well as some properties of the regions (kinematic distances and luminosities), are given in Table \ref{table0}.


\section{ Observational Results}
\label{sec.3}

The five regions observed were previously reported by \citet{Sanchez2011} using lower angular resolutions of approximately 9\arcsec and 6\arcsec at 3.6 and 1.3 cm, respectively, which are roughly six times lower than our resolution of 0.9\arcsec. With our higher resolution observations, we were able to resolve several sources into multiple components.

We detected continuum emission at 2 and 6 cm toward all regions and confirmed the previous detections at 3.6 and 1.3 cm, (see Figure \ref{Figure1}). We find that two of the candidates for HC HII reported by \citet{Sanchez2011} are multiple systems (G53.04+0.11 and G53.14+0.07), and we made a new detection of one source (G53.04+0.11nw). The remaining three HC HII regions are single-component sources (G34.82+0.35, G40.28$-$0.22 and G48.99$-$0.30). We detected a total of 10 sources in the five observed regions, which are very compact ($\lesssim$ 20 mpc) and mostly appear as unresolved sources. In addition, their flux densities are mostly between 0.1 and 1 mJy, except for G48.99$-$0.20, which is about $\sim$10 times stronger.

In Table \ref{par-o} we report the peak positions (from the highest angular resolution 6 cm images), the integrated flux densities (at all four wavelengths), and the measured source sizes (at 6 cm) defined as half the geometric mean of the major and minor axes, which were determined by fitting 2-D Gaussians with CASA. We used 6 cm images for these fits because they have better resolution and $uv$-coverage than the 2 cm images. 
The flux densities were measured with the IMFIT task in CASA, which fits two-dimensional Gaussians to the image data. Most of the sources were unresolved, so the integrated flux obtained is equivalent to that of a point source. In the few cases with indications of resolved emission, IMFIT consistently recovered the total flux.
We also report in Table \ref{par-o}  the observed brightness temperature and the spectral index obtained from a linear least squares fit to S$_{\nu} \propto \nu^{\alpha}$ between 1.3 and 6 cm.  In most cases the spectral index is consistent with thermal free-free emission from (partially optically thick) HII regions or, alternatively,  collimated/spherical winds. We discuss both scenarios below.

\section{Models and Methods applied}
\label{sec.4}

\subsection{HII regions}

The evolution of HII regions is a complex process that develops in clouds of varying densities. These regions may start in cores of constant density and spread toward areas where the density decreases following a power law. During the formation and expansion of HII regions, there is a critical exponent that plays a crucial role, as it determines whether the cloud will be completely ionized or not \citep{Franco1990}. If the power-law exponent exceeds 1.5, the HII region may enter the so-called champagne phase, characterized by supersonic expansion. Furthermore, in disk-shaped clouds, this evolution can lead to the appearance of neutral, high-velocity outflows.\citep{Franco1989}.

To interpret the radio spectra of small HII regions, we use models from \citet{Olnon1975} to estimate the physical parameters for  spherical, cylindrical, and Gaussian geometries, using the equations in appendix \ref{appe}. We fit the flux density as a function of frequency by adjusting the free parameters of the equations which are the radius, R, and the electron density, n$_{e}$, for an isothermal 10$^4$ K plasma.

We found that the results from the three models were very similar for each source, with the parameters R and n$_{e}$ predicted by the models falling within the errors of the fit (Figure \ref{esp1}). For simplicity, we present the results for the spherical distribution case only. In addition, we report the emission measure, optical depth, ionized mass, and minimum Lyman continuum photon flux obtained from the spherical models. The ionized mass was calculated assuming a homogeneous sphere of ionized hydrogen of radius $R$ and electron density $n_e$, using $M_{\text{i}} = \tfrac{4}{3}\pi R^3 n_e m_p$. The minimum Lyman continuum photon flux was derived by equating ionizations and recombinations in the same volume, with $N_{\text{Ly}} = \tfrac{4}{3}\pi R^3 n_e^2 \alpha_B$, where $\alpha_B$ is the case B recombination coefficient for $10^4$~K gas.  Varying the electron temperature between 8,000 and 11,000 K produces about a 10\% change in radius values and a 20\% change in electron density. Table \ref{par-aj} shows the results of the fitting process.

We note that the Strömgren sphere model assumes a uniform density structure. The spherical and cylindrical cases in the Olnon models likewise adopt uniform density, which explains the small differences between them. Since our sources were not spatially resolved and the power-law models require more observational constraints, we adopted the spherical distribution as the most practical approach.

\subsection{Spherical Stellar Wind}

According to an alternative scenario proposed by \citet{Rodriguez2012}, the free-free emission may originate from a spherical (isotropic) stellar wind. Equation (\ref{flux}) relates the flux density to the frequency, mass-loss rate, terminal velocity $V_{\infty}$ and distance. The equation is attributed to \citet{Panagia1975} and \citet{Felli1981}, and assumes a temperature of 10$^{4}$ K for the ionized gas and a mean atomic weight per electron of 1.2.

\begin{align}\label{flux}
\left[\frac{S_{\nu}}{\mathrm{mJy}}\right] &=
1.28 \left[\frac{\nu}{\mathrm{GHz}}\right]^{0.6}
\left[\frac{\dot{M}}{10^{-5}\, M_{\sun}\,\mathrm{yr}^{-1}}\right]^{4/3} \nonumber \\
&\times \left[\frac{V_{\infty}}{10^{3}\,\mathrm{km\,s}^{-1}}\right]^{-4/3}
\left[\frac{D}{\mathrm{kpc}}\right]^{-2}
\end{align}

Adopting a typical terminal velocity of 10$^{3}$ km s$^{-1}$ \citep{Rodriguez1983} and using the 6 cm flux densities reported in Table \ref{par-o}, we can calculate the corresponding mass loss rates. These values are reported in column 2 of Table \ref{wind}.

Additionally, the minimum number of Lyman-continuum photons needed to ionize the envelope $N_{c}$ was derived by \citet{Felli1981} and is given by equation \ref{ion}.

\begin{align}\label{ion}
\left[\frac{N_{c}}{\mathrm{s}^{-1}}\right] &=
2.9\times 10^{47}
\left[\frac{\dot{M}}{10^{-5}\, M_{\sun}\,\mathrm{yr}^{-1}}\right]^{2}
\left[\frac{V_{\infty}}{10^{3}\,\mathrm{km\,s}^{-1}}\right]^{-2} \nonumber \\
&\times \left[\frac{R}{10\, R_{\sun}}\right]^{-1}
\end{align}

To estimate the maximum mass loss rate for zero-age main-sequence (ZAMS) stars, we use equation \ref{ion}, replacing N$_c$ with the closest value of ionizing photon rate (N$_i$) and radius (R) from \citet{Panagia1973} and adopt a terminal velocity of 1000 km~s$^{-1}$.  The resulting mass loss rates are shown in column 5 of Table \ref{wind}. We also included the luminosity from \citet{Panagia1973} and the infrared luminosity from Table \ref{table0} in Table \ref{wind}.  Except for two sources, G48.99$-$0.30 and G53.04+0.11se, the IR luminosity from Table \ref{table0} and the luminosity of a similar ZAMS star are within a factor of 3 of one another.  Thus, except for these two cases, a spherical stellar wind is a plausable model.  For the two exceptions, the large luminosity excess reported in Table \ref{wind} cannot be explained by the corresponding stellar spectral type.

\subsection{Collimated Stellar Wind (Jet)}

An alternative to the spherical wind model is the case of a collimated wind or jet. In this case, the mass loss rate can be estimated using the results of \citet{Reynolds1986}.

\begin{eqnarray}\label{csw}
\left[\frac{\dot{M}}{10^{-6}\, M_{\sun}\,\mathrm{yr}^{-1}}\right]
& = & 0.938\, \chi_{0}^{-1}
\left[\frac{\mu}{m_{p}}\right]
\left[\frac{V_{w}}{10^{3}\,\mathrm{km\,s}^{-1}}\right]
\left[\frac{S_{\nu}}{\mathrm{mJy}}\right]^{3/4} \nonumber\\
& \times & \left[\frac{\nu}{10\,\mathrm{GHz}}\right]^{-3\alpha_{op}/4}
\left[\frac{D}{\mathrm{kpc}}\right]^{3/2} \nonumber\\
& \times & \left[\frac{\nu_{m}}{10\,\mathrm{GHz}}\right]^{-0.45+3\alpha_{op}/4}
\left[\tan \theta_{0}\right]^{3/4} \nonumber\\
& \times & \left[\frac{T}{10^{4}\,\mathrm{K}}\right]^{-0.075}
\left[\sin i\right]^{-1/4} \nonumber\\
& \times & \left[\frac{2.1\,(\alpha_{op}-1.3)}{(\alpha_{op}-2)(\alpha_{op}+0.1)}\right]^{-3/4}
\end{eqnarray}

Here, $\chi_{0}=1$ represents the ionized fraction of hydrogen, $\mu$ is mean particle mass per hydrogen atom, $m_{p}$ is the proton mass and is assumed to be $\frac{\mu}{m_{p}}=1$. $V_{w}$ is the collimated wind velocity, $S_{\nu}$ is density flux at $\nu$=5 GHz, T=10$^4$ K, D is the distance to region from Table \ref{table0}, $\nu_m$ is the turnover frequency and  $\alpha_{op}$ is the opaque spectral index. The turnover frequencies were taken as 10 GHz for all sources, except for G40.28$-$0.22 where we used 20 GHz. The opaque spectral index $\alpha_{op}$ was determined between 6 and 8 GHz. However, for the sources G34.82+0.35 and 40.28$-$0.22, the opaque spectral index was taken from Table \ref{par-o}  because the fluxes at 8 GHz appear to be affected by extended emission, due to the larger beam size at that frequency. We calculate the jet mass loss rate for two (arbitrarily chosen but reasonable) cases: one with $i$=45\textdegree and $\theta_0$=27\textdegree, the other with  $i$=25\textdegree and $\theta_0$=10\textdegree. Here $i$ represents the inclination angle of the jet with respect to the sky while $\theta_0$ is the jet opening angle. These mass loss rates are presented in Table \ref{jet}.

\section{NOTES ON INDIVIDUAL SOURCES}
\label{sec.5}

In this section we present the detailed results and discussion for the five individual fields we observed. The sources are all massive young stellar objects embedded in molecular clumps, whose temperatures were determined by ammonia observations of \citet{Cyganowski2013} and masses from ATLASGAL continuum observations of \citet{Urquhart2011}. The clump temperatures range from 19 to 31 K, while the clump masses range from 215 to 12300 M$_{\sun}$ (Table \ref{table6}).  For each source we comment on the HII region model and the two stellar wind model results.

\subsection{G34.82+0.35}

The G34.82+0.35 region, also known as IRAS 18511+0146, Mol75 and RAFGL 5542, was proposed as a candidate precursor to a UC HII region \citep{Molinari1996}. \citet{Vig2007} presented an extensive analysis and suggested that IRAS 18511+0146 is a protocluster with the most massive member being a precursor to a Herbig Ae/Be star. They presented 450 and 850 $\mu$m images of a dusty molecular region about 1\arcmin in size that showed evidence of internal structure. Their MIPS images at 70 and 24 $\mu$m resolved the region into three distinct sources. The brighter two of these, A and B, were considered active star formation sites, owing to the presence of water and methanol masers in their vicinity (\citet[22 GHz]{Urquhart2011}, \citet[44 GHz]{Kurtz2004}, \citet[6.7 GHz]{Pandian2009} and \citet[6.7 GHz]{Fontani2010}).  In addition, \citet{Zhang2005} report a molecular outflow in the region, which they detected in the CO(2-1) line.

We find two radio continuum sources: G34.82+0.35w, corresponding to Source A from \citet{Vig2007}, and G34.82+0.35e, corresponding to their Source B (see Fig. \ref{Figure1}). The angular separation between the two is 21\arcsec, corresponding to a projected distance of 0.37 pc. G34.82+0.35w (source A), was detected at 3.6 cm by \citet{Watt1999}.  Their image also shows weak 3.6 cm emission at the position of G34.82+0.35e (source B) but they do not report it as a detection.  In our observations, both sources are unresolved at 2 cm and marginally resolved at 6 cm.  G34.82+0.35e was detected by \citet{Sanchez2011} at 1.3 and 3.6 cm, while G34.82+0.35w was marginally detected at 3.6 cm and not detected at 1.3 cm. Both sources are shown in Figure \ref{Figure1}.

The salient aspect of the two sources is that G34.82+0.35w is IR-loud and radio-quiet, while G34.82+0.35e is the reverse; see \citet{Vig2007}.  Both sources are embedded within the same molecular cloud, which both \citet{Vig2007} and \citet{Watt1999} estimate to have a mass of order 10$^3$~M$_\sun$; the former based on 450/850~$\mu$m observations, the latter on CO observations.
Thus, the two sources appear to be distinct massive star formation sites within a single molecular core; their different behavior in terms of radio versus infrared emission perhaps indicating different stages of development.

\citet{Vig2007} estimate that G34.82+0.35e is ionized by a B2-B1 star, with a Lyman photon flux of $\sim 1.3 \times 10^{45}$~s$^{-1}$, in close agreement with our spherical HII region estimate of $1.0 \times 10^{45}$~s$^{-1}$.  Their bolometric luminosity of 4400 L$_\sun$ is also in good agreement with our HII region model estimate of 5000 L$_\sun$, also indicative of a B1 type star.  This close agreement suggests that a single star may be exciting the G34.82+0.35e region.
That G35.82+0.35w is weaker in radio, but substantially brighter in the infrared, suggests that multiple protostars, of lower mass, are powering this region.

Considering the spectral indices of 0.4 and 0.6, the small sizes of both sources, and their luminosities obtained via the ionizing photon rate (0.29 and $0.50 \times 10^{4}$~L$_{\sun}$)  which are less than the total infrared luminosity of the region ($2 \times 10^{4}$~L$_{\sun}$), we consider that these sources may be small and weak UC HII regions.  However, the radio continuum properties are not exclusive of stellar winds or jets, and we cannot rule out these possibilities.

\subsection{G40.28$-$0.22}

G40.28$-$0.22 is associated with IRAS 19031+0621, which is embedded in an infrared dark cloud (IRDC) \citep{Peretto2009, Cyganowski2008} and has been classified as an Extended Green Object (EGO) (see Figure \ref{Figure2}) and a candidate Massive Young Stellar Object (MYSO) \citep{Cyganowski2008}.  Both the Bolocam Galactic Plane Survey \citep{Rosolowsky2010} and the ATLASGAL Survey \citep{Csengeri2014} report a dusty cloud, surrounding the radio continuum source.  Class I and II methanol masers and water masers are reported by \citet{Chen2011}, \citet{Pandian2009}, and \citet{Cyganowski2013}, respectively.   \citet{Pandian2010} detected 1.3 cm and 6.9 mm continuum emission coincident with the 6.7 GHz methanol maser.  Their results suggest that the 6.9 mm emission has a strong contribution from dust, thus we do not include this point in our modeling of the radio free-free emission.  

\citet{Sanchez2011} report a single unresolved source at 3.6 cm and 1.3 cm with an estimated size between 1 and 10 mpc and an emission measure of $\gtrsim$10$^{9}$ pc cm$^{-6}$.  We detect the continuum source at both wavelengths (6 and 2 cm) as shown in Figure \ref{Figure1}, where we also show the re-processed 1.3 and 3.6 cm images. In Figure \ref{Figure2} we show the infrared emission, exhibiting the EGO characteristics.

The fit results  (Figure \ref{esp1},  Table \ref{par-aj}), indicate that the source is small, with a radius of $\sim$ 66 AU (0.32 mpc) with a high density (2.4$\times$10$^{6}$ cm$^{-3}$) and an emission measure of 3.7$\times$10$^{9}$ pc cm$^{-6}$.  The Lyman photon flux corresponds to a single ZAMS B1 star.
\citet{Ge2014} reported that the EGO cloud core has a size of 52 mpc, a mass of 25 M$_{\sun}$, and a density of 5.53$\times$10$^{6}$ cm$^{-3}$.  We note the general agreement between the reported molecular core density and the ionized gas density.

In summary,  considering the  spectral index of 1.64 (between 5 and 22 GHz) as well as the small size,  high density, and emission measure,  the source meets the usual criteria for a HC HII region \citep{Kurtz2002, Kurtz2005}.  If the source were a stellar wind the radio emission would imply a mass loss rate of about $10^{-5}$ M$_{\sun}$ year$^{-1}$ corresponding to a stellar luminosity of 3.8$\times$10$^{4}$ L$_{\sun}$ (see Table \ref{wind}), nearly a factor of 2 higher than the region's reported IR luminosity of 2.13$\times$10$^{4}$ L$_{\sun}$.

To our knowledge, there is no evidence in the literature for the presence of an outflow in this region, and the radio continuum morphology has yet to be spatially resolved.  We consider this source to be a prime candidate for detailed studies of HC HII regions.

\subsection{G48.99-0.30}
\label{Dis.G48}

This region is part of the W51B complex \citep{Koo1999} and is associated with IRAS 19201+1400. A continuum source was reported by the CORNISH survey \citep{Hoare2012} at 6 cm, coincident with IR emission. The \citet{Sanchez2011} data show a point source to the north of a possible bright rim (see Fig. 1). Although several surveys in the infrared and radio continuum have included this region (RMS, CORNISH, ATLASGAL, and HI-GAL), the compact source has not been well-studied. Water \citep{Nagayama2015} and methanol \citep{Pandian2009} masers were found near the centimeter source.  \citep{Nagayama2015} report a distance of 5.62$^{+0.59}_{-0.49}$ kpc based on VERA  parallax observations of the water masers.

As noted by \citet{Sanchez2011}, extended emission is present in the field. To suppress the resulting imaging artifacts, we used only visibilities from baselines longer than 20 k$\lambda$, corresponding to a largest angular size of 6$''$.  We detected the point source, G48.99$-$0.30, at 6 and 2 cm wavelengths, as shown in Figure \ref{Figure1} (along with the re-processed data at 1.3 and 3.6 cm).  An infrared image is shown in Figure \ref{Figure2}, where the bright rim is clearly visible.  The source is associated with the most massive (12,000 M$_{\sun}$) and hottest (27.6 K) clump among the four detected within the IRDC by \citet{Traficante2015}.

Our fitting procedure for an HII region (Table \ref{par-aj}) indicates that the source is quite small (3.7 mpc), of high density (2.3$\times$10$^{5}$ cm$^{-3}$) and emission measure (3$\times$10$^{8}$ pc cm$^{-6}$). These results and the spectral index (+0.48) suggest that the source could be a HC HII region. The Lyman photon flux could be provided by a single ZAMS-type B0.5 star, which would have a bolometric luminosity of about $1.1 \times 10^{4}$L$_{\sun}$ --- somewhat lower than the IR luminosity of $4.5 \times 10^{4}$L$_{\sun}$ reported by \citet{Maud2015}.

Our analysis of the stellar wind scenario (Table \ref{wind}) indicates that this model is not viable because the stellar bolometric luminosity inferred from the ionizing photon rate would be 15$\times$ higher than the IR luminosity of the region. Similarly, the jet scenario (Table \ref{jet}) is unlikely because the predicted mass-loss rate of 3.9$\times$10$^{-4}$ M$_{\sun}$ year$^{-1}$ is extremely high.

\subsection{G53.04+0.11}

This region is associated to IRAS 19266+1745, located at 9.4 kpc (Table \ref{table0}) with an IR luminosity of 5$\times$10$^{4}$ L$_{\sun}$ \citep{Lu2014}. It is associated with a massive clump detected at submillimeter wavelengths \citep{Urquhart2014,  Williams2004} with a mass  between 1300 and 2800 M$_{\sun}$. A massive outflow, seen in CO, was reported by \citet{Beuther2002b}.

Only one continuum source was reported by \citet{Sanchez2011},  detected at both 3.6 cm and 1.3 cm with flux densities of 1.8 and 2.5 mJy, respectively. The RMS survey detected this source at 6 cm with a flux density of 1.1 mJy (beam $\sim$ 0.7\arcsec) \citep{Urquhart2009} and it was marginally detected (3-4 $\sigma$) in the 6~cm CORNISH survey \citep{Hoare2012}. Weak water, class II and class I methanol masers were reported by \citet{Sridharan2002}, \citet{Pandian2009} and \citet{Litovchenko2011}.

After re-calibrating the data at 3.6 and 1.3 cm, we detect two sources (Figure \ref{Figure1}). The more southern source is brighter, and coincides with the source reported by \citet{Sanchez2011} and \citet{Urquhart2009}. The more northern (and western) source is fainter and is detected at 3.6 cm but not at 1.3 cm.

In our new observations at 6 and 2 cm we detect both the northern and southern sources, and we resolve the southern source into two components.  We label these three as
G53.04+0.11nw, G53.04+0.11se and G53.04+0.11sw (see figure \ref{Figure1}). The separation between  G53.04+0.11nw and the southern sources is $\sim$9.5\arcsec while G53.04+0.11sw is found 2.4\arcsec  west of  G53.04+0.11se.  The latter is the brightest of the three sources while G53.04+0.11sw is the faintest. All three sources are unresolved at our resolution; their flux densities are listed in Table \ref{par-o}.

G53.04+0.11se at 6 cm and 2 cm corresponds to the  1.3  and 3.6 cm continuum  source at reported by \citet{Sanchez2011}, but with a lower angular resolution (7\arcsec). With our better resolution of 0.9\arcsec and 3\arcsec at 2 and 6 cm, respectively, we resolve this source into two components which we label as se and sw.  In addition, with our better sensitivity we detect the nw source, which is also evident in the re-imaged data at 3.6 cm.  Flux densities for the three components are reported in Table 2.

The sw and se sources lie at the edge of extended IR emission as seen in the GLIMPSE data.  The nw source is coincident with an IR point source, also seen in the GLIMPSE images (see Fig. 2).

More recently, IRAS 19266+1745 was observed by \citet{Rosero2016} at 6 and 1.3~cm and by \citet{Tatiana2023} at 7~mm.
Our sources G53.04+0.11se and G53.04+0.11sw are associated with the continuum sources A and B, respectively, as reported by \citet{Rosero2016}.  Their source C is located further to the south and is not reported by us, while our source G53.04+0.11nw is not reported by them.  \citet{Tatiana2023} report 7 mm continuum emission in sources C1, C3, and C5 (corresponding to se, sw and nw in our nomenclature).  In all three cases, their flux densities appear to be dominated by dust emission; we do not consider them in our analysis.

We obtained flat to positive spectral indices for the three sources G53.04+0.11se ($\alpha$=0.12), G53.04+0.11sw ($\alpha$=0.71), and G53.04+0.11nw ($\alpha$=0.42), which are consistent with thermal emission in HII regions. However, this is inconsistent with the findings of \citet{Rosero2016} for their sources A and B (se and sw in our nomenclature).  Although our 6 and 3.6 cm fluxes are in reasonable agreement with their two 6 cm values (at 4.9 and 7.4 GHz), our 2 and 1.3 cm fluxes are notably higher than their two 1.3 cm values (at 20.9 and 25.5 GHz).  As a result, we obtain substantially more positive spectral indices than they report. Possible reasons for this discrepancy include differences in the data reduction and variability in the region. Because both sources show a similar increase in flux density with respect to the \citet{Rosero2016} values, we suspect that differing data reduction procedures are the more likely cause.

Our observations and spherical modelling suggest that all three sources are very small (2-6 mpc from model fits). The densities and emission measures are of similar order of magnitude (10$^{5}$ cm$^{-3}$ and 10$^{8}$ pc cm$^{-6}$) and the ionizing star of each is likely a B0.5-type ZAMS star, assuming that each ionized region has only  one embedded star.  Although the size is quite small, the density is not really high enough to consider these as hypercompact HII regions.  If the photoionized Str\"omgren sphere model holds, then the physical interpretation would be that these are small and rather weak 
UC HII regions.

G53.04+0.11sw could be a stellar wind with a mass loss rate of $2.2 \times 10^{-5}$ M$_{\sun}$ yr$^{-1}$, corresponding to a luminosity of $6.5 \times 10^{4}$ L$_{\sun}$. This is slightly larger than the luminosity of the region ($5 \times 10^{4}$ L$_{\sun}$). We cannot conclusively distinguish between a stellar wind and a UC HII region because the luminosity of the region is roughly consistent with both scenarios.

It is unlikely that G53.04+0.11se is a stellar wind (Table \ref{wind}), as the luminosity  corresponding to its mass loss rate appears is much higher than the luminosity of the region (39.8 versus $5 \times 10^4$~L$_\sun$). Additionally, it is unlikely that these three sources are jets because there was no detection of SiO(1-0) emission \citep{Zapata2009}, which is a tracer of molecular flows. We note that the IR luminosity reported in Tables 1 and 5 is based on {\it IRAS} data, and hence is the total luminosity of the region, not that of the individual components.

\subsection{G53.14+0.07}

We detected three continuum sources (G53.14+0.07se, G53.14+0.07sw, G53.14+0.07n) at 2 cm. They are arranged in a triangle, with G53.14+0.07n located at the apex, about 4\arcsec north of the other two sources (see Figure \ref{Figure1}). We did not detect the southern sources at 6 cm with an rms of 40 $\mu$Jy beam$^{-1}$, and they were not detected by \citet{Sanchez2011} at 3.6 cm, who report a single source at 1.3 cm.  Our re-reduced image of their 1.3 cm data shows extended, diffuse emission to the south of G53.14+0.07n, which may have a contribution from the two southern sources.
At 2 cm, G53.14+0.07sw is slightly elongated in the east-west direction and has the highest flux density ($S_{2 cm} = 0.3$ mJy) of the three sources (see Table \ref{par-o}). G53.14+0.07se and G53.14+0.07n are unresolved and have flux densities of approximately 0.16 mJy.

All three sources we detect are coincident with the MM1 core of the G053.11+0.05 IRDC reported by \citet{Rathborne2006}.  MM1 is the most massive core within the IRDC, with a mass of 124 M$_\sun$ based on 1.2~mm observations.  The source is associated with infrared emission, class I and II methanol masers, and water masers (see Figure \ref{Figure2}  and Table 4).

G53.14+0.07n coincides with YSO1 reported by \citet{Kim2018}, while G53.14+0.07sw is within several arc seconds of their position for YSO2.  They report a molecular outflow, detected in H$_2$, and suggest that one of the two YSOs is the driving source.  The elongated morphology of G53.14+0.07sw, although not well-aligned with the outflow, suggests that it may be the outflow driver.

We could not use the Olnon models for these sources as for the other regions, owing to a reliable flux density at only one frequency. However, we were able to estimate an upper limit for their size of about 10 mpc.
We can then apply the standard equations for a spherical, homogeneous, isothermal HII region (with an assumed
electron temperature of $10^4$~K) and also for the spherical wind scenario, described in Section 4.2.

The calculated properties of the three sources are consistent with small UC HII regions, whose Lyman continuum photon flux ($\sim 10^{43}$  s$^{-1}$) corresponds to a ZAMS B3 or B2 star \citep{Panagia1973}.
More massive stars are possible, because the sizes reported in Table 2 are upper limits. These regions might also be stellar winds.  The luminosities (Table \ref{wind}) would be slightly higher than that of the region, but only by factors of a few, which is within the uncertainty.

\section{Discussion and conclusions}
\label{sec.6}

The simplest model for a HC HII region is a Str\"omgren sphere within a very high density medium and with an abundant supply of ionizing photons. 
However, as several decades of research have shown, and as alluded to in the introduction and Section 4.1, the simplicity of this model fails to address numerous aspects of these regions.   Mass infall and outflow, along with circumstellar accretion disks, substantially complicate the dynamics of HII regions; this is just as much the case for hypercompact regions as it is for their larger and lower density cousins, the compact and ultracompact HII regions.

The observed source sizes from the radio continuum for our sample of candidate HC HII regions are quite small, even for hypercompact regions (see Table 2).  Nevertheless, the electron densities (assuming the spherical Olnon model, see Table 3), although fairly high for UC HII regions, are somewhat on the low side for HC HII regions.  Perhaps most notable from our modeling is that --- at these size scales --- cylindrical, spherical and Gaussian distributions predict very similar continuum spectra.  It would appear that only highly-resolved images will be able to distinguish the true density profile.

We investigated five candidate HC HII regions that were previously observed at 1.3 and 3.6 cm. To expand on those observations, we conducted new observations at 2 and 6 cm using the VLA and detected a total of 10 compact sources in these five regions. Of these sources, five are newly detected, as three of the original sources were resolved into multiple components.
The observed source radii are in the range of 4 to 10 mpc. Using the spherical Olnon model and the source flux density distributions, the model radii are smaller, ranging from 0.4 to 3.7 mpc.  Using a spherical stellar wind model, the mass loss rates range from 0.17 to $15.9 \times 10^{-5}$~M$_\sun$~yr$^{-1}$.

Of the 10 sources we report,  we favor the interpretation of a HC HII region for two of them, G40.28$-$0.22 and G48.99$-$0.30.   These sources are particularly good candidates for further HC HII region studies.
For the remaining eight sources we consider small, weak UC HII regions to be the more likely interpretation.
For five of these (the two components of G34.82+0.35 and the three components of G53.14+0.07) we cannot rule out either the spherical wind or the jet interpretation.  
One of our top candidates for a jet is G34.82+0.35w. Its spectral index (+0.6) and mass loss rate (between 5 and $9 \times 10^{-6}$~M$_\sun$~yr$^{-1}$) are reaonable for a jet. In addition, it is associated with an outflow reported in CO (J=2-1) \citep{Zhang2005} and methanol masers that may have formed in gas shocked by the flows.
For the remaining three sources, of G53.04+0.11, the sw source could be a wind, but the se source is unlikely to be, based on luminosity arguments.  We reject the jet hypothesis for all three sources, based on the lack of shocked gas tracers.

When compared to the size of our solar system, given by the heliopause of about 130 AU or 0.6 mpc, we note that our regions are similar in size to the solar system. In contrast, the stellar density in the core of the Orion Trapezium is approximately $2 \times 10^{4}$ stars pc$^{-3}$ \citep{Hillenbrand1998}, equivalent to having one star per 40 mpc assuming a uniform stellar distribution in the core. Hence, we believe that the HC HII regions host only one or two stars.

In summary, our observations and modeling indicate that most sources are best classified as small, weak UC HII regions, although a few remain viable candidates for HC HII regions. One source, G34.82+0.35w, is a strong jet candidate.  Future multi-wavelength observations with higher angular resolution will be essential to resolve these sources and confirm their nature.

\section*{Acknowledgements}

We thank the anonymous referee for helpful suggestions and comments that improved the quality of the paper.

A.S.-M. acknowledges support from the RyC2021-032892-I grant funded by \linebreak[4] \nolinkurl{MCIN/AEI/10.13039/501100011033} and by the European Union `Next GenerationEU'/PRTR, as well as the program Unidad de Excelencia Mar\'\i{}a de Maeztu CEX2020-001058-M, and support from the PID2023-146675NB-I00 (MCI-AEI-FEDER, UE).

T.R.E. is grateful to The World Academy of Science (TWAS) for a postdoctoral grant during which most of this work was completed.

\clearpage


\begin{figure*}
\centering
\begin{subfigure}{0.45\textwidth}
  \includegraphics[width=\textwidth]{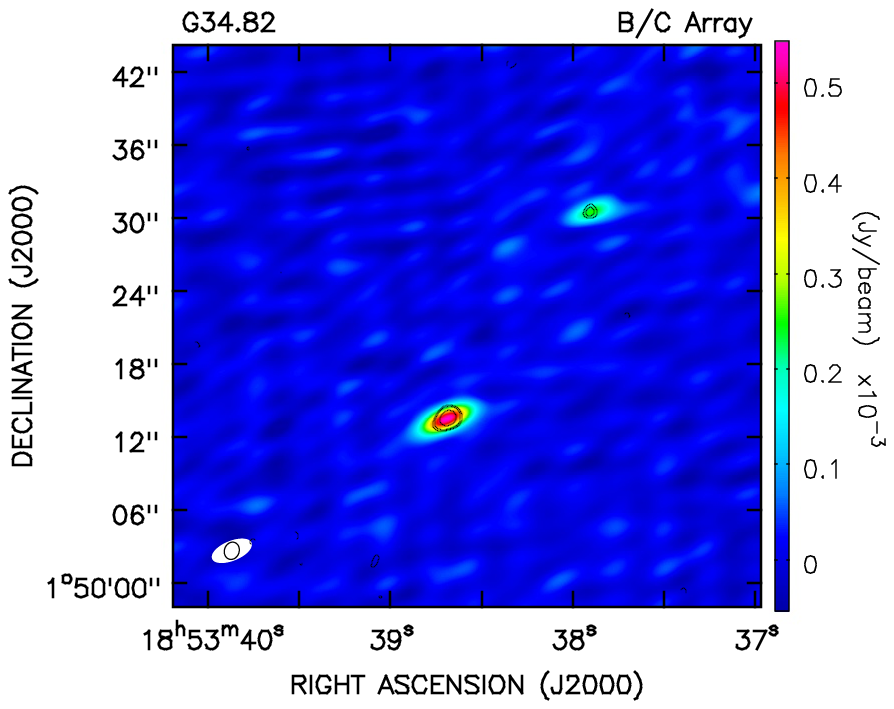}
  \vspace*{-1.5cm}
  \subcaption{G34.82+0.35 (2 y 6 cm)}
\end{subfigure}
\hfill
\begin{subfigure}{0.45\textwidth}
  \includegraphics[width=\textwidth]{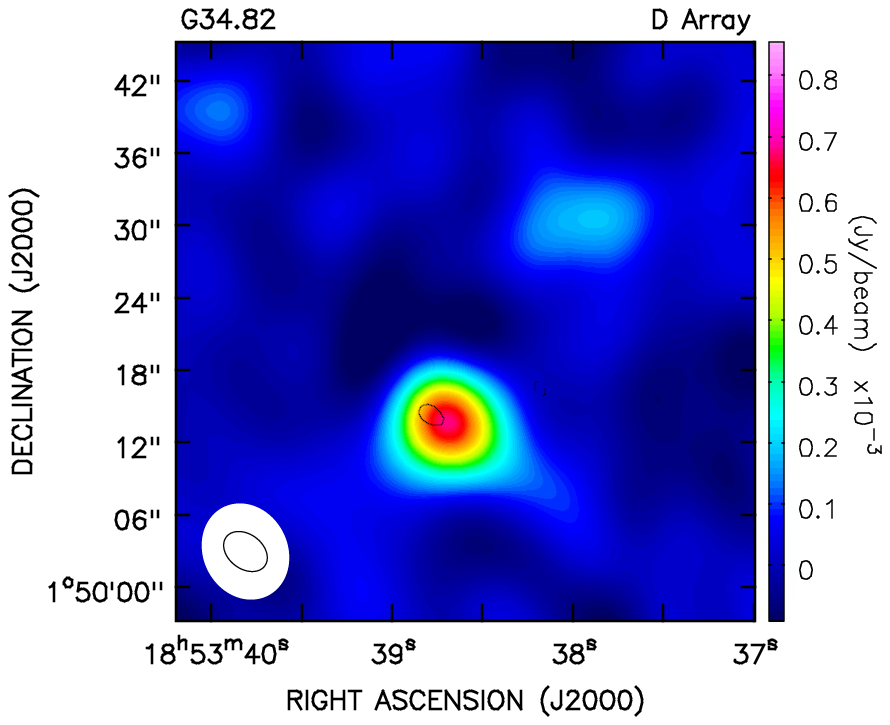}
  \vspace*{-1.5cm}
  \subcaption{G34.82+0.35 (1.3 y 3.6 cm)}
\end{subfigure}

\begin{subfigure}{0.45\textwidth}
  \includegraphics[width=\textwidth]{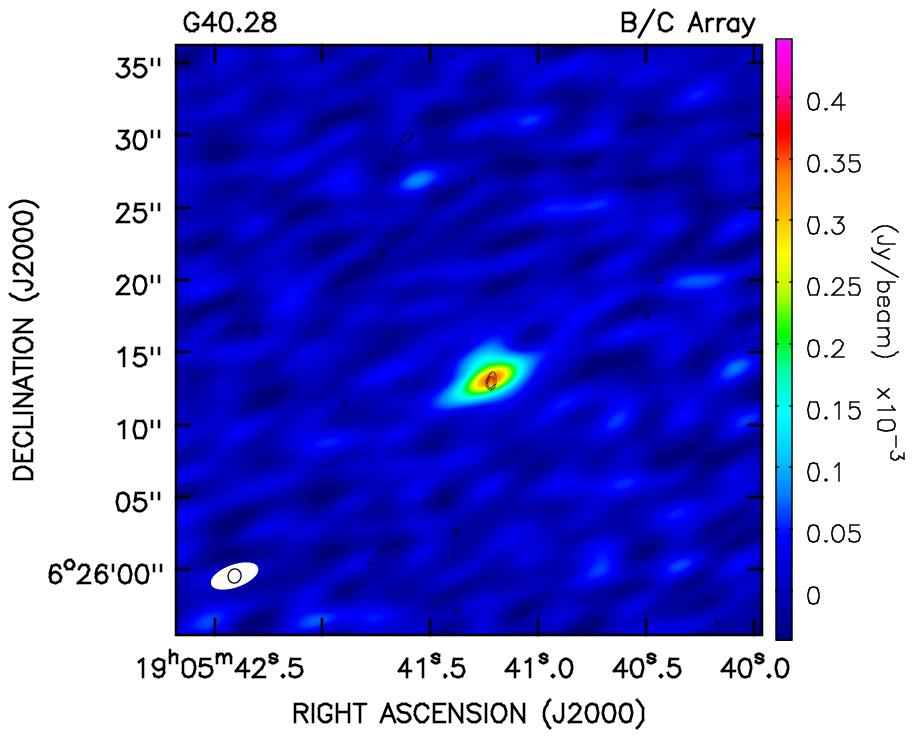}
  \vspace*{-1.5cm}
  \subcaption{G40.28$-$0.22 (2 y 6 cm)}
\end{subfigure}
\hfill
\begin{subfigure}{0.45\textwidth}
  \includegraphics[width=\textwidth]{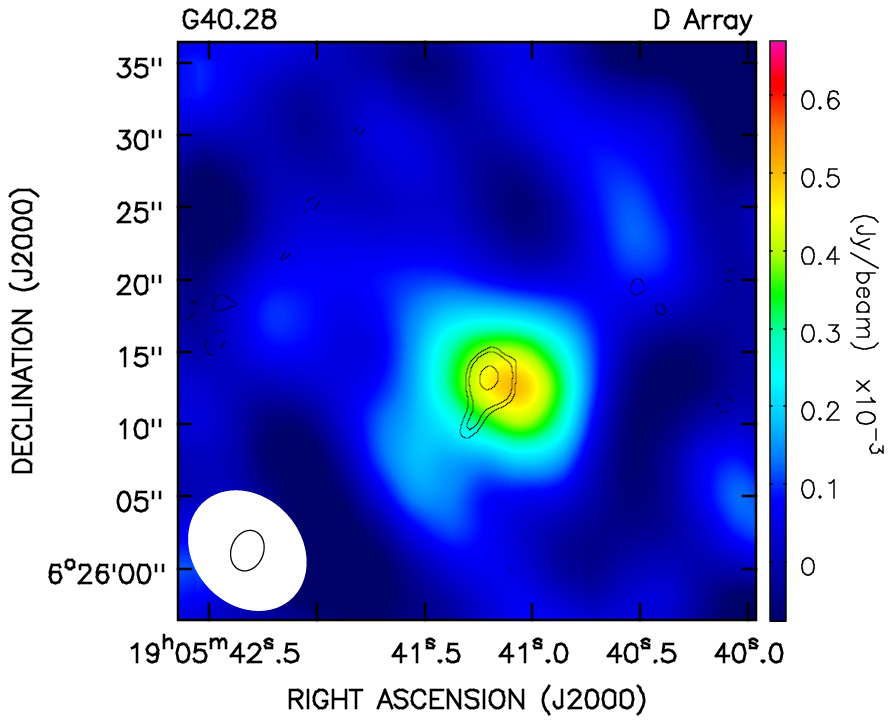}
  \vspace*{-1.5cm}
  \subcaption{G40.28$-$0.22 (1.3 y 3.6 cm)}
\end{subfigure}

\caption{Radio continuum images. Left: color map of the 2~cm emission (VLA-C) overlaid with contours of the 6~cm emission (VLA-B). Right: color map of the 3.6~cm emission (VLA-D) overlaid with contours of the 1.3~cm emission (VLA-D). The contour levels are $-$3, 3, $2^{n}$...32, $n$=2...5 of the map rms given in Table \ref{table0}. The synthesized beams are shown in the bottom-left corner of each map.}
\label{Figure1}
\end{figure*}

\begin{figure*}\ContinuedFloat
\centering

\begin{subfigure}{0.45\textwidth}
  \includegraphics[width=\textwidth]{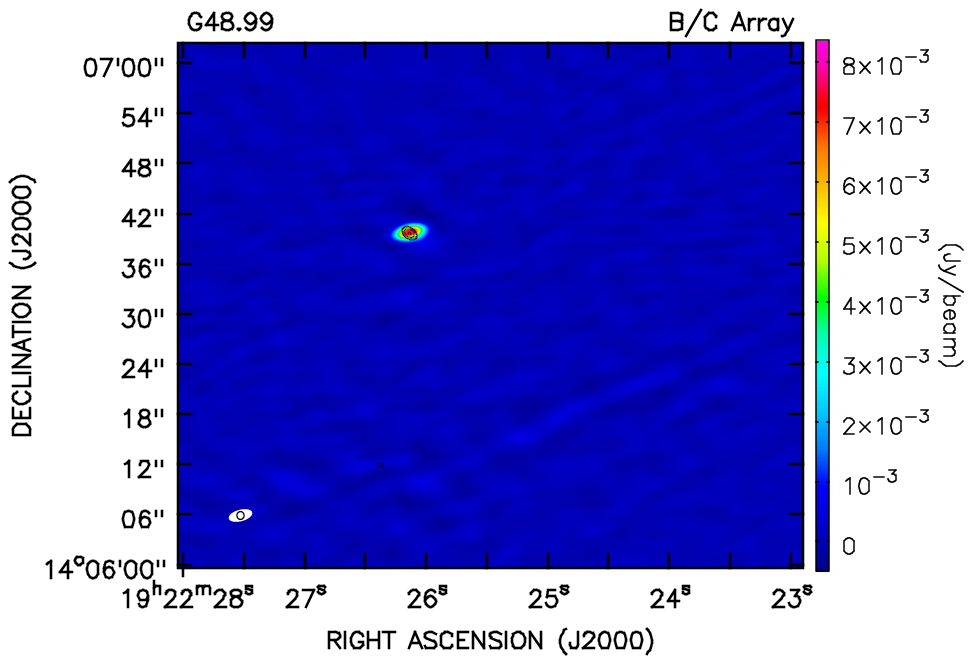}
  \vspace*{-1.5cm}
  \subcaption{G48.99$-$0.30 (2 y 6 cm)}
\end{subfigure}
\hfill
\begin{subfigure}{0.45\textwidth}
  \includegraphics[width=\textwidth]{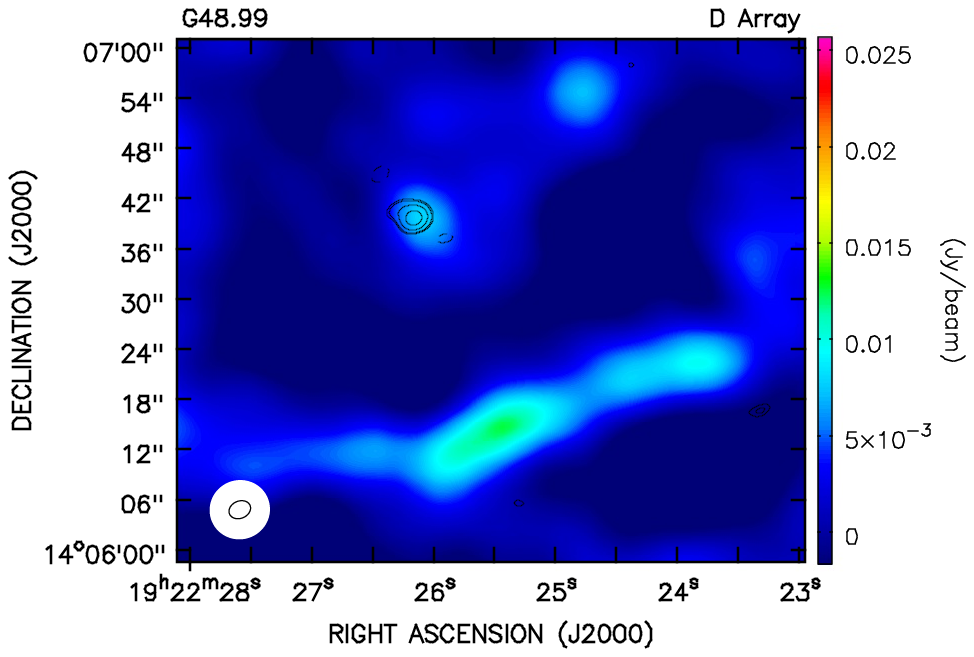}
  \vspace*{-1.5cm}
  \subcaption{G48.99$-$0.30 (1.3 y 3.6 cm)}
\end{subfigure}

\begin{subfigure}{0.45\textwidth}
  \includegraphics[width=\textwidth]{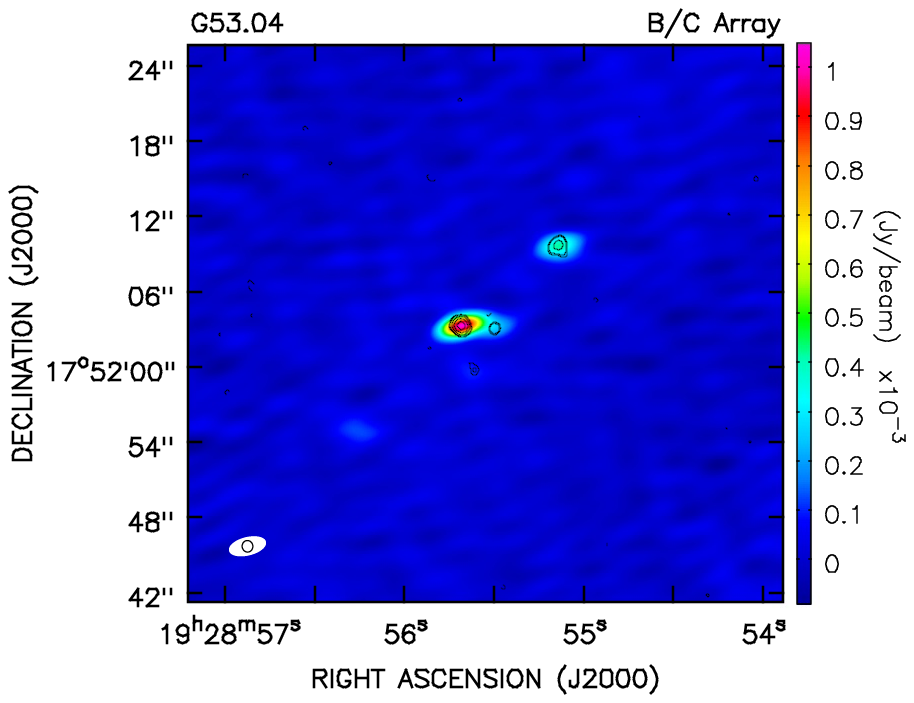}
  \vspace*{-1.5cm}
  \subcaption{G53.04+0.11 (2 y 6 cm)}
\end{subfigure}
\hfill
\begin{subfigure}{0.45\textwidth}
  \includegraphics[width=\textwidth]{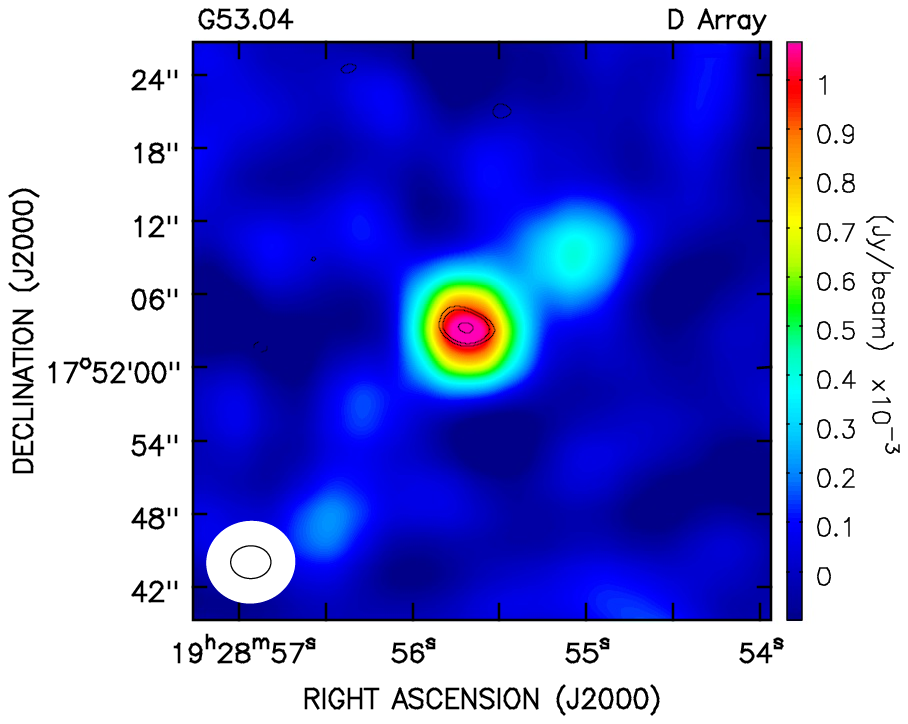}
  \vspace*{-1.5cm}
  \subcaption{G53.04+0.11 (1.3 y 3.6 cm)}
\end{subfigure}

\begin{subfigure}{0.45\textwidth}
  \includegraphics[width=\textwidth]{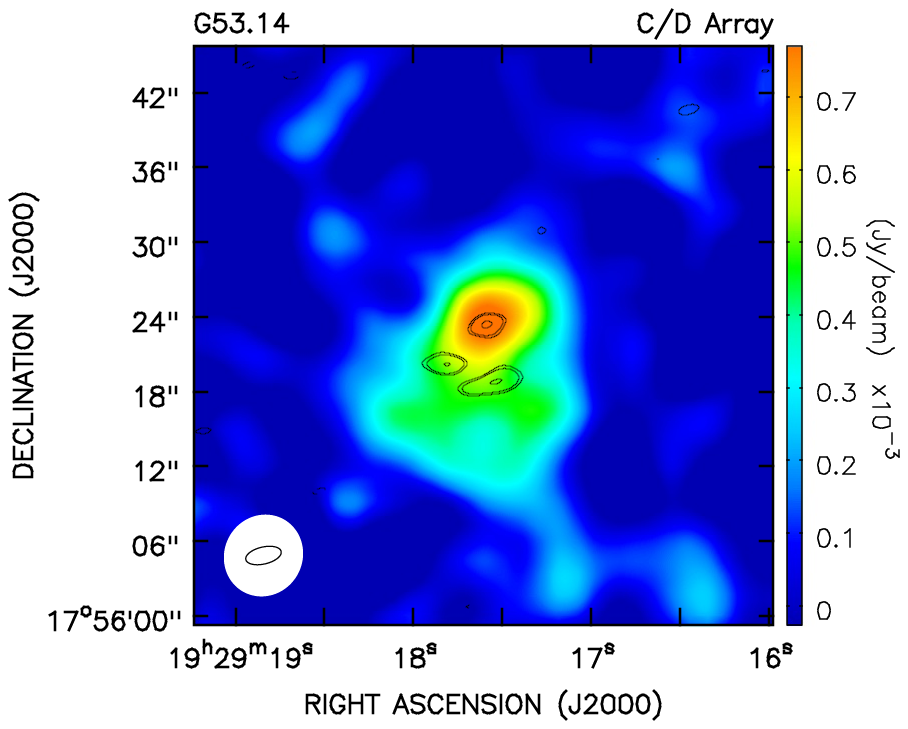}
  \vspace*{-1.5cm}
  \subcaption{G53.14+0.07 (1.3 y 2 cm)}
\end{subfigure}

\caption{Cont.Radio continuum images. For G53.14+0.07, the 1.3 cm emission (VLA-D) color image is overlaid with a contour map of the 2 cm emission (VLA-C).}
\label{Figure1}
\end{figure*}

\begin{figure*}[ht!]
\centering

\begin{subfigure}{0.48\textwidth}
  \centering
  \includegraphics[trim=2cm 0cm 4cm 3cm, clip, width=\textwidth]{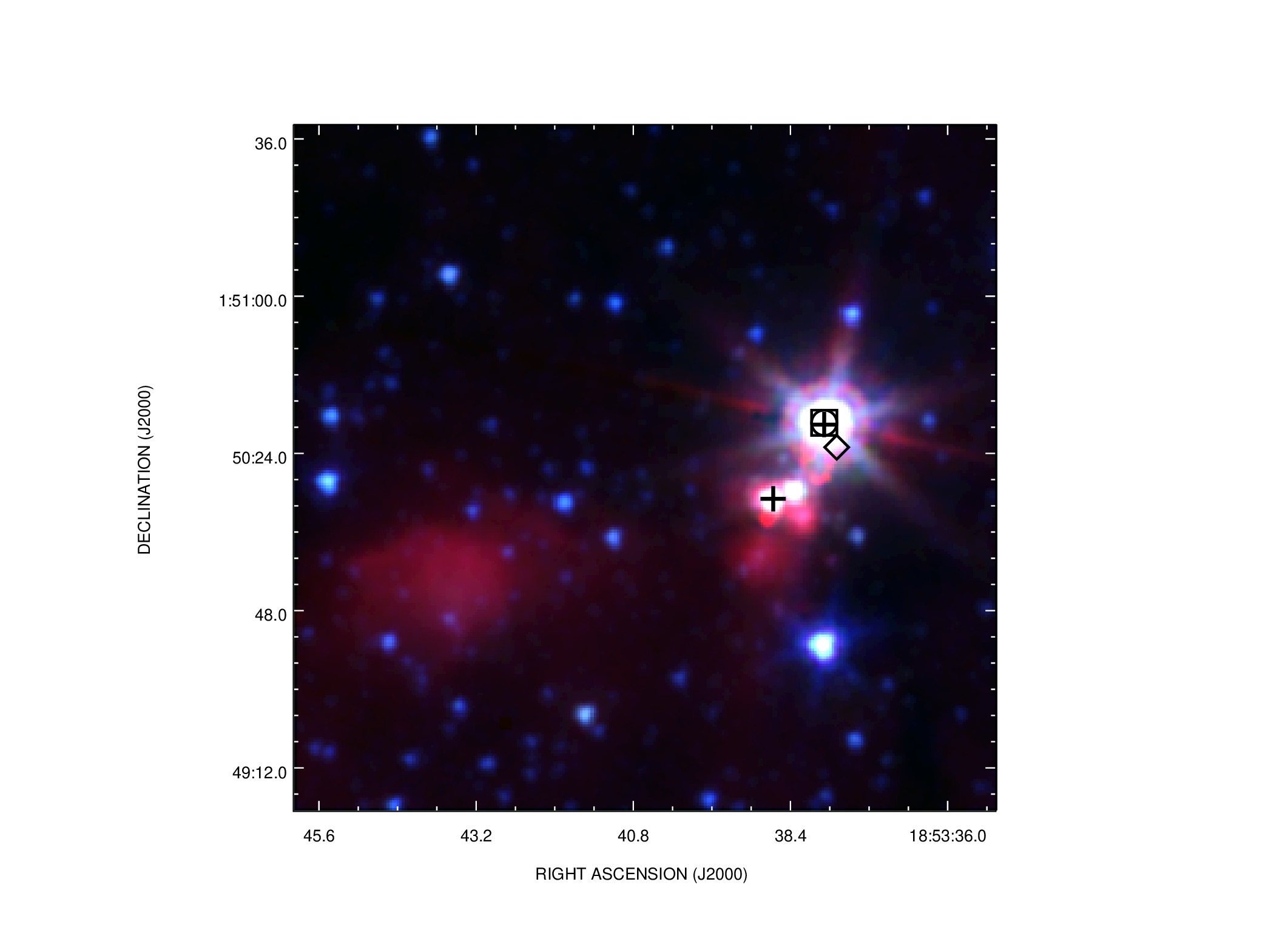}
  \subcaption{G34.82+0.35}
\end{subfigure}
\hfill
\begin{subfigure}{0.48\textwidth}
  \centering
  \includegraphics[trim=2cm 0cm 3cm 3cm, clip, width=\textwidth]{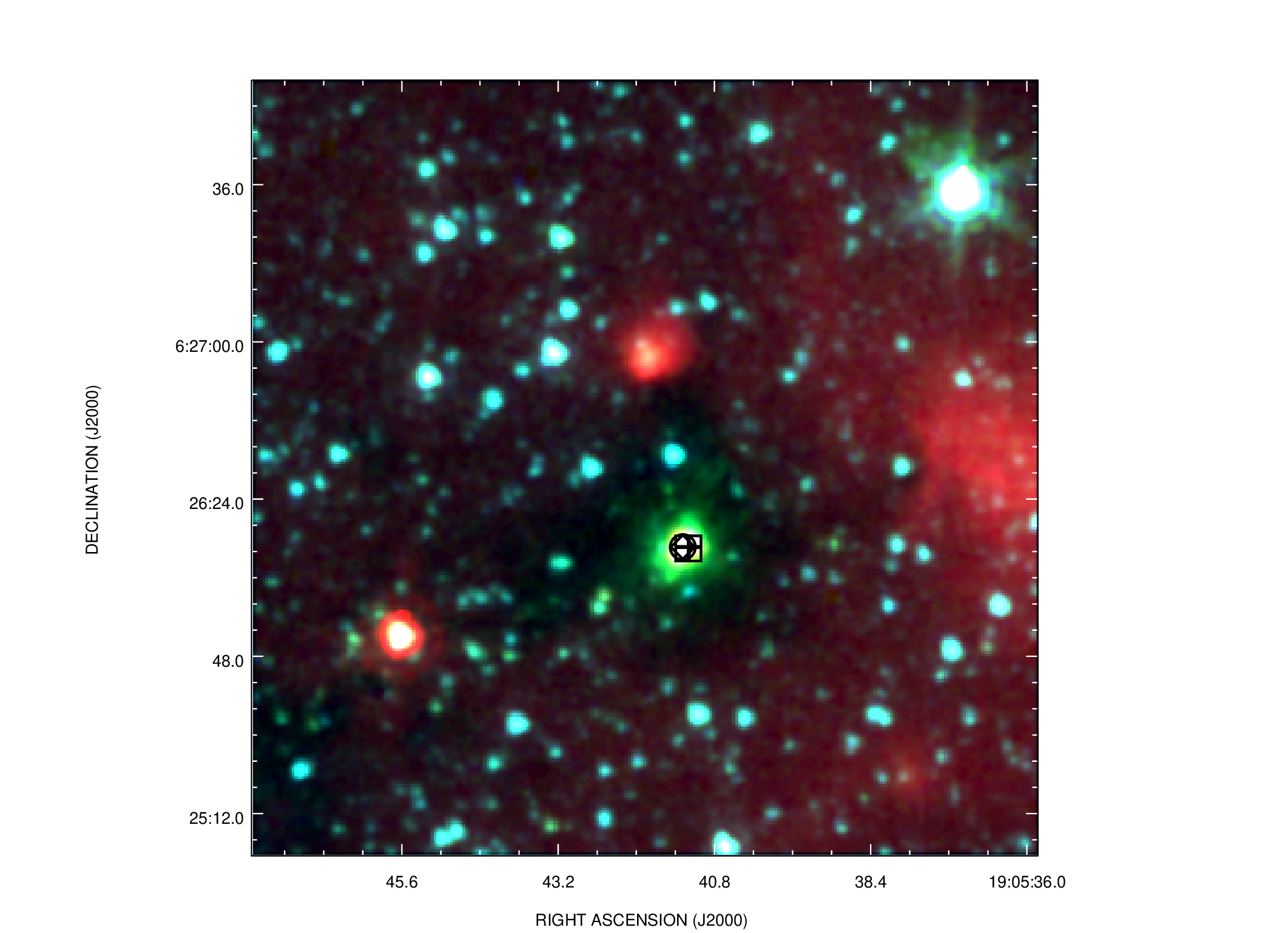}
  \subcaption{G40.28$-$0.22}
\end{subfigure}

\vspace{2mm}

\begin{subfigure}{0.48\textwidth}
  \centering
  \includegraphics[trim=0cm 0cm 4cm 2cm, clip, width=\textwidth]{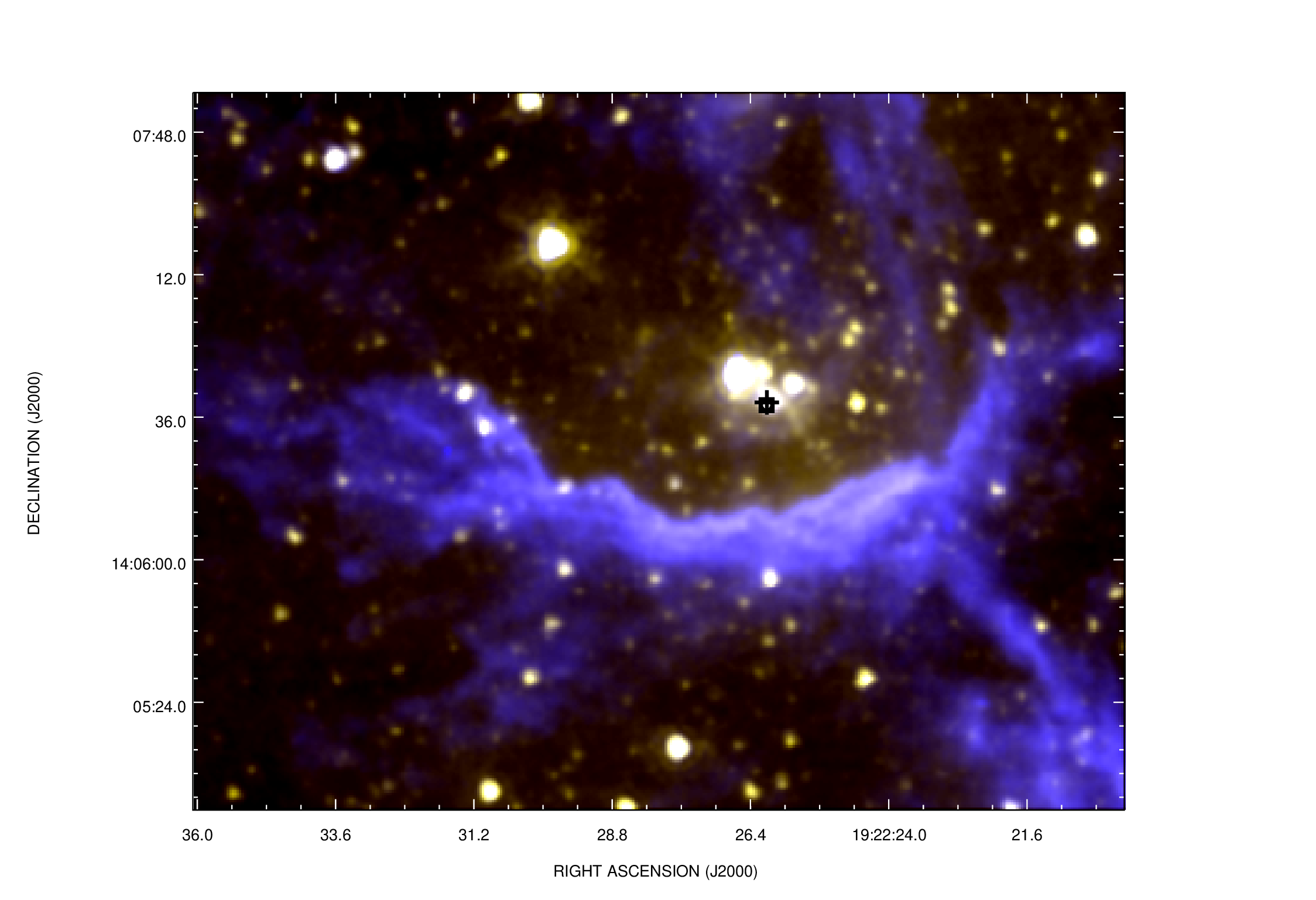}
  \subcaption{G48.99$-$0.30}
\end{subfigure}
\hfill
\begin{subfigure}{0.48\textwidth}
  \centering
  \includegraphics[trim=0cm 0cm 3cm 2cm, clip, width=\textwidth]{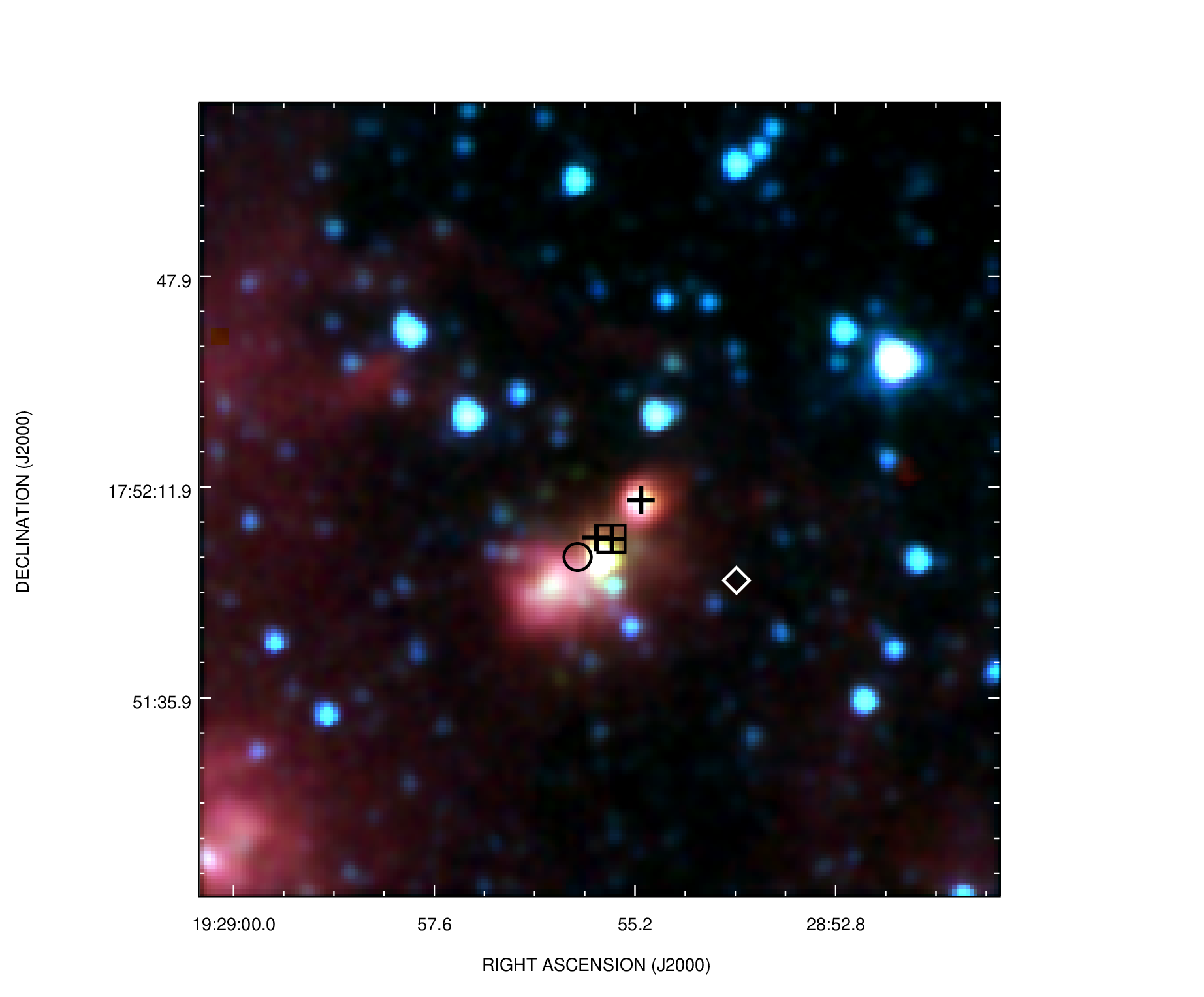}
  \subcaption{G53.04+0.11}
\end{subfigure}

\vspace{2mm}

\begin{subfigure}{0.48\textwidth}
  \centering
  \includegraphics[trim=0cm 0cm 2cm 2cm, clip, width=\textwidth]{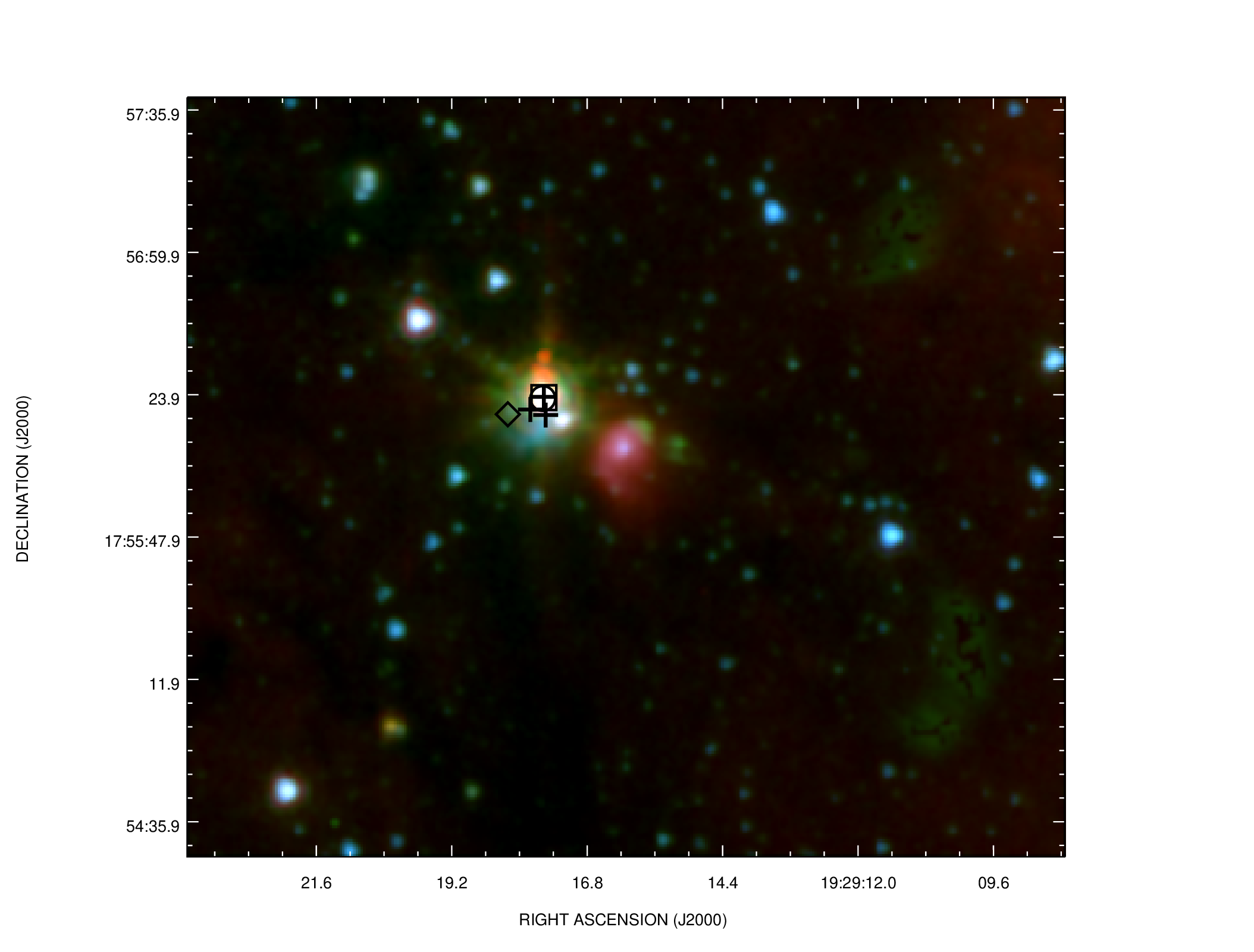}
  \subcaption{G53.14+0.07}
\end{subfigure}

\caption{Overlay of infrared emission at 8.0 $\mu$m (red), 4.5 $\mu$m (green), and 3.6 $\mu$m (blue) taken from the GLIMPSE IRAC images \citep{Churchwell2001, Benjamin2003}. The radio sources we detect are marked with crosses. The positions of class~I and class~II methanol masers are indicated by diamonds and squares, respectively, and the positions of water masers are indicated by circles. References for each tracer are listed in Table~\ref{tracers}.}
\label{Figure2}
\end{figure*}

\begin{figure*}[ht!]
\centering

\includegraphics[width=0.32\textwidth]{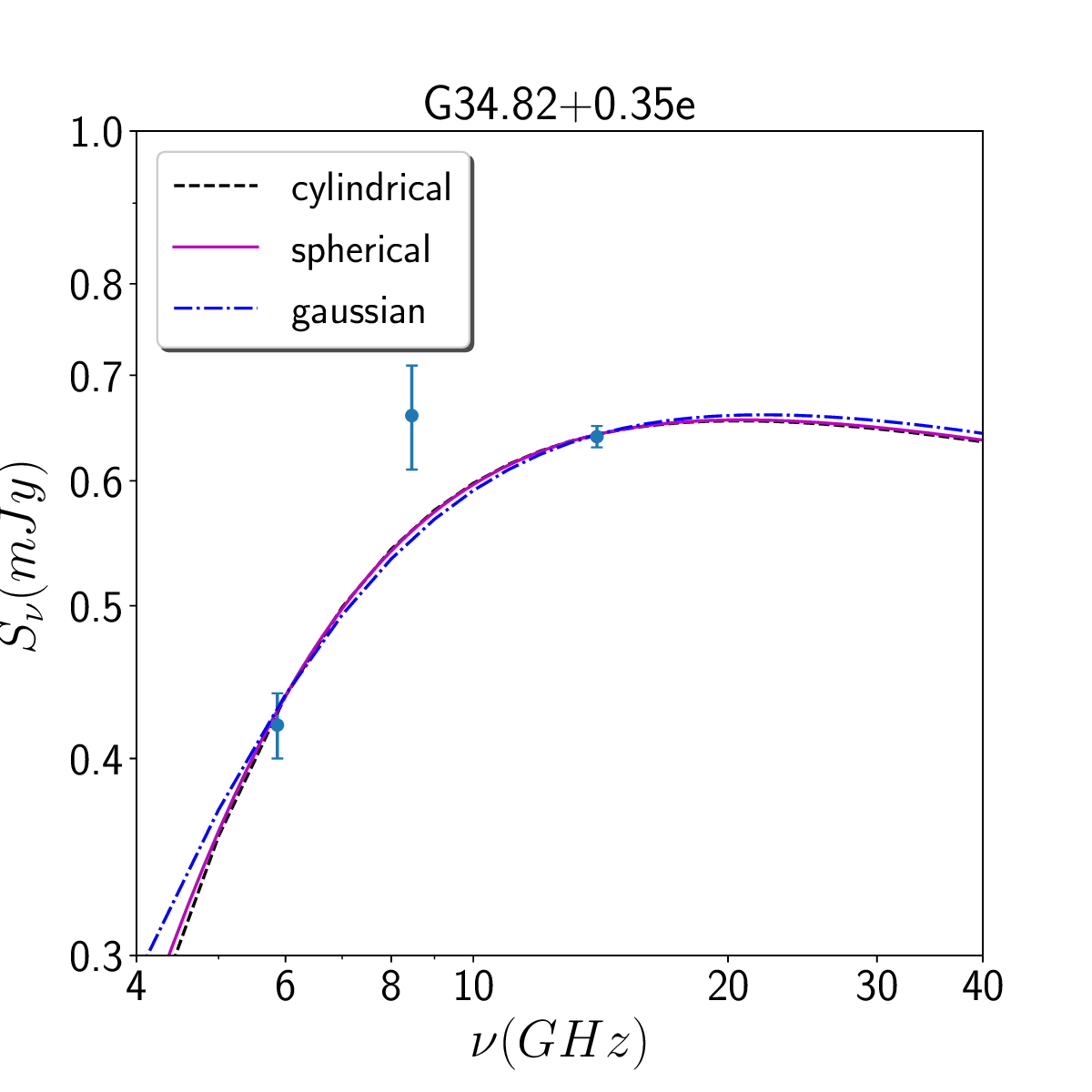}
\includegraphics[width=0.32\textwidth]{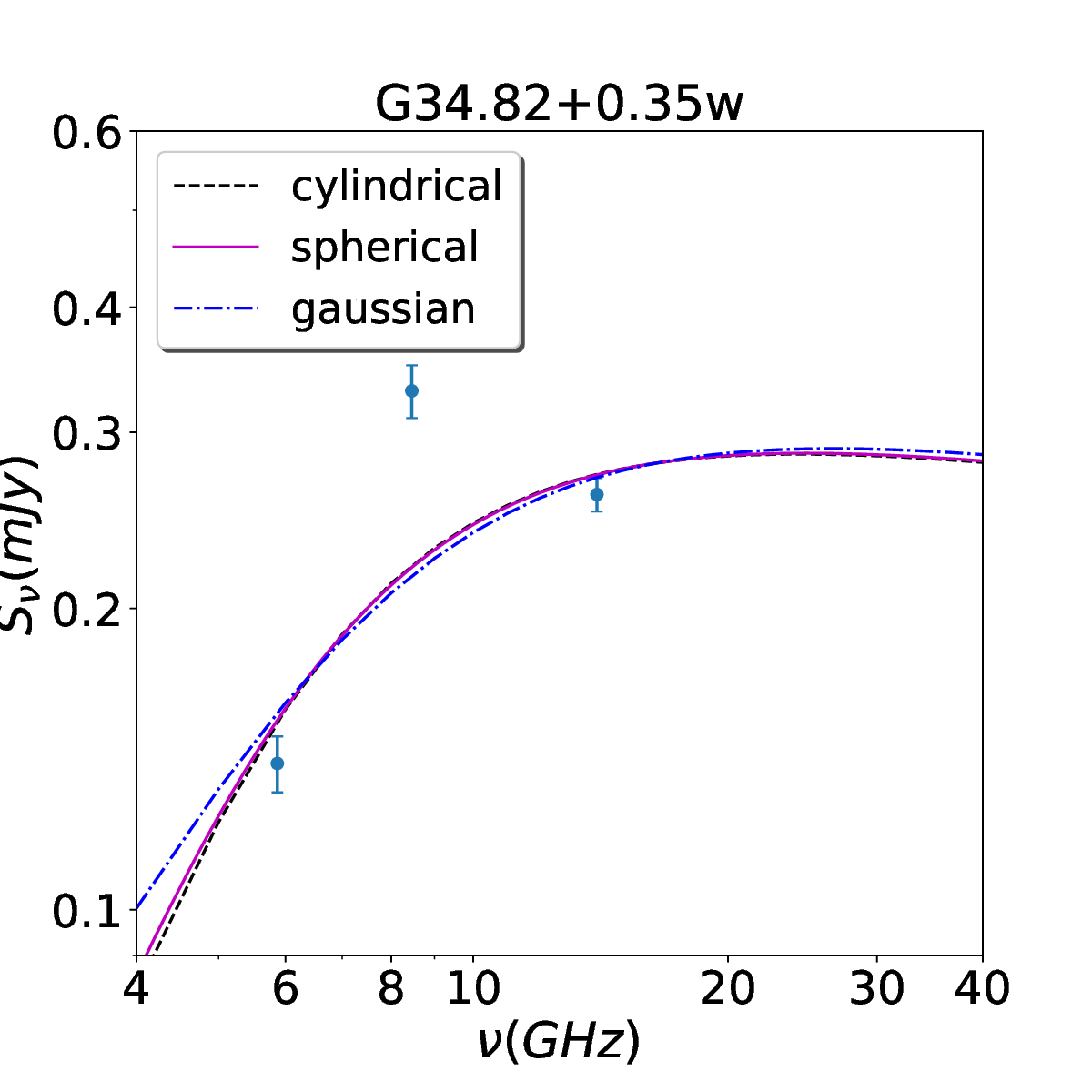}
\includegraphics[width=0.32\textwidth]{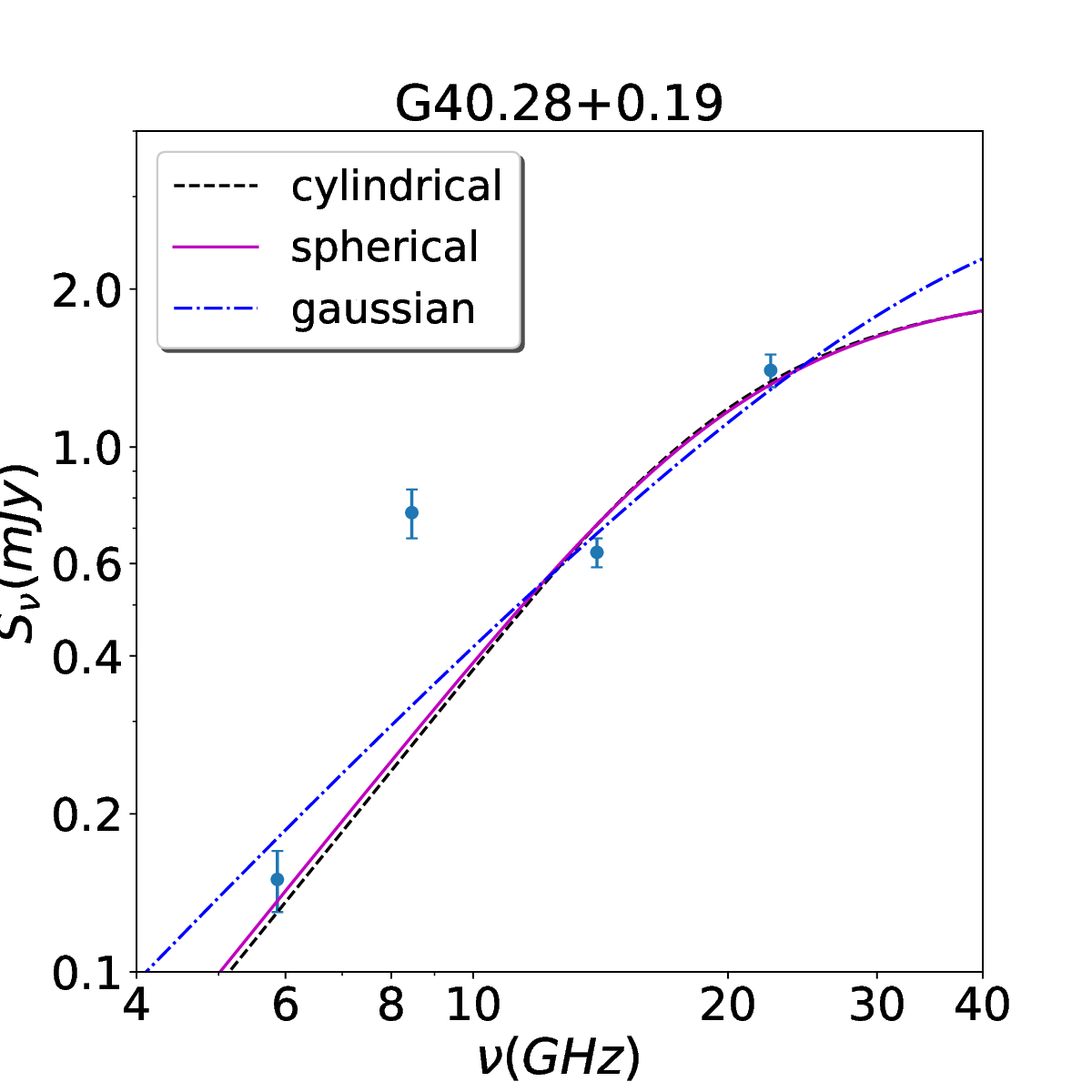}

\vspace{2mm}

\includegraphics[width=0.32\textwidth]{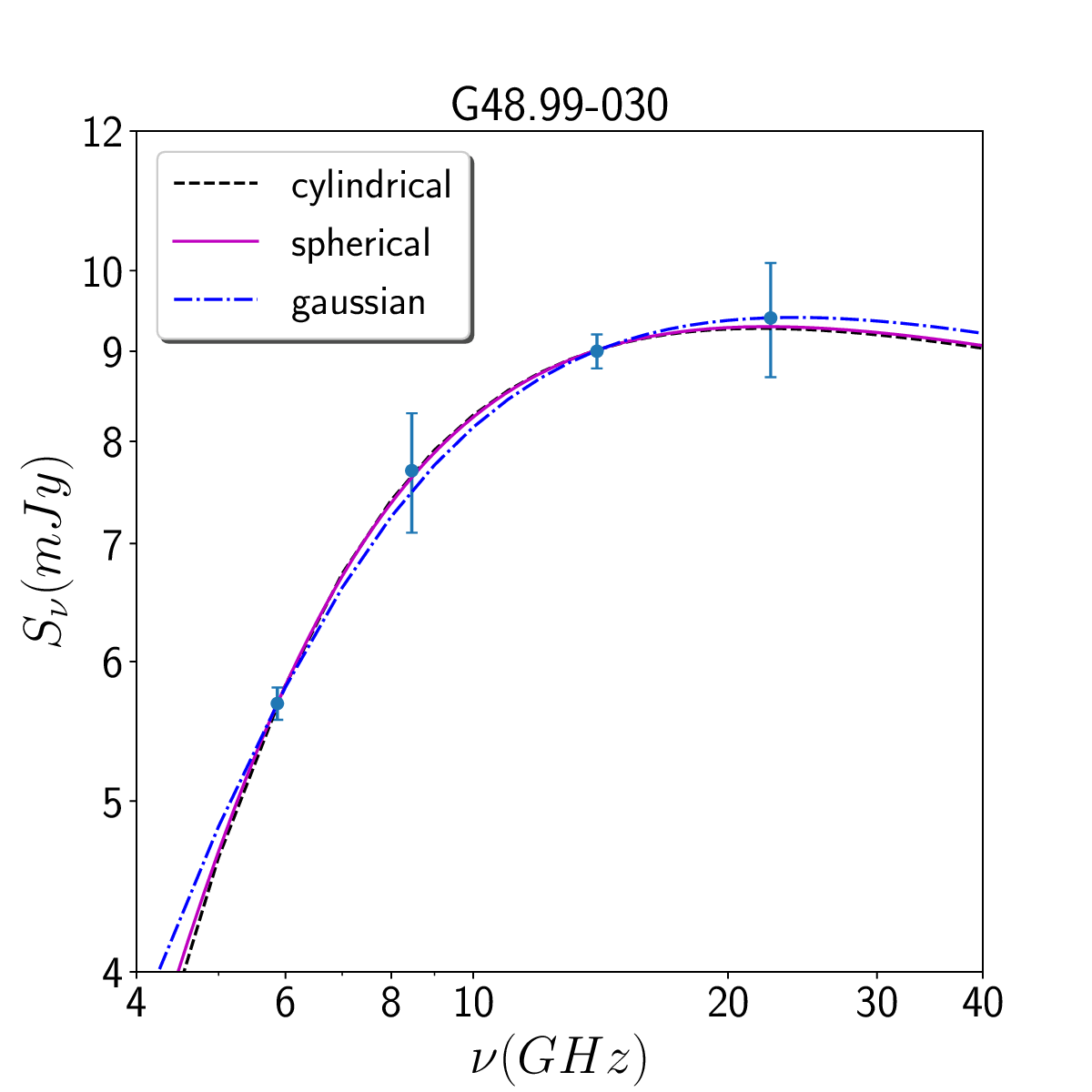}
\includegraphics[width=0.32\textwidth]{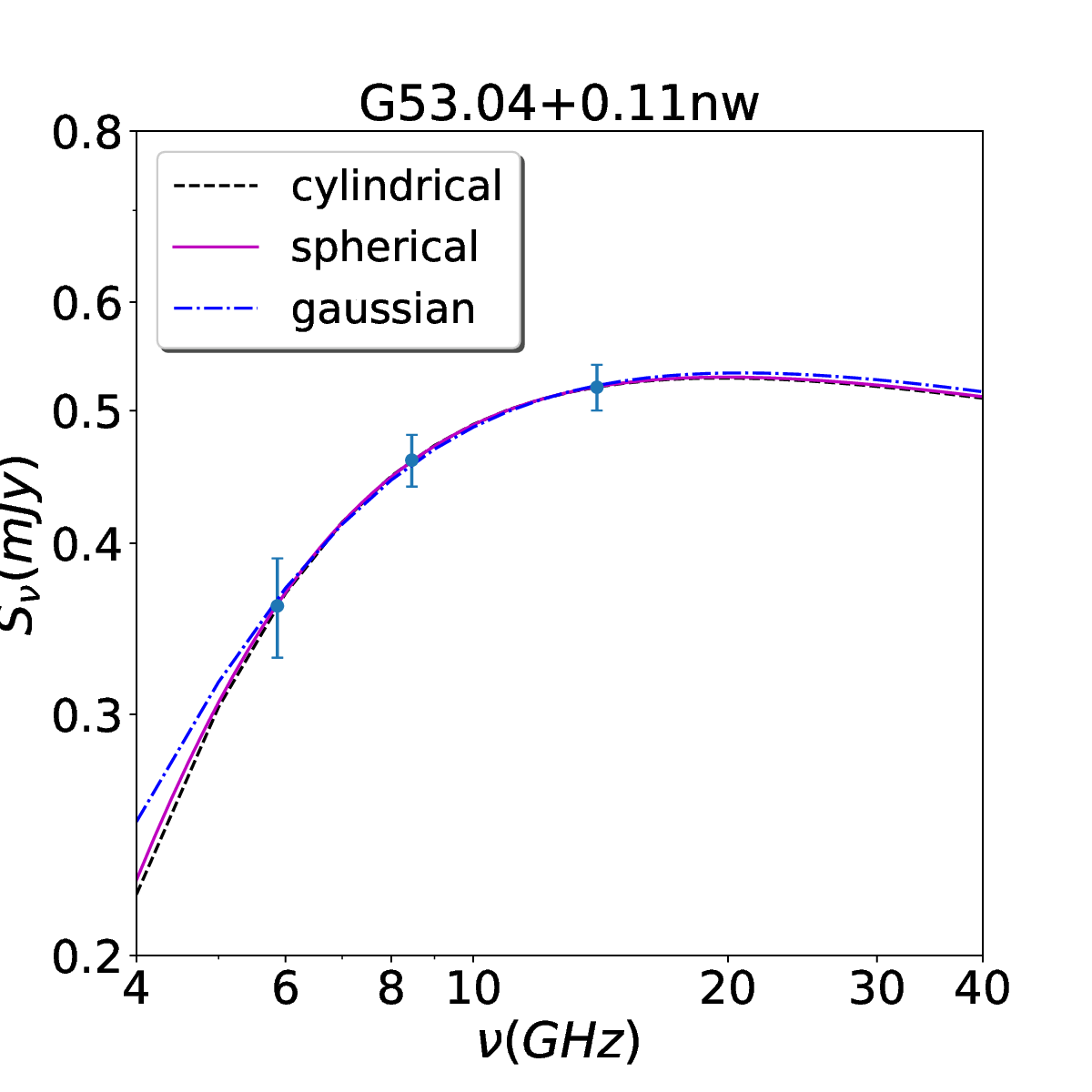}
\includegraphics[width=0.32\textwidth]{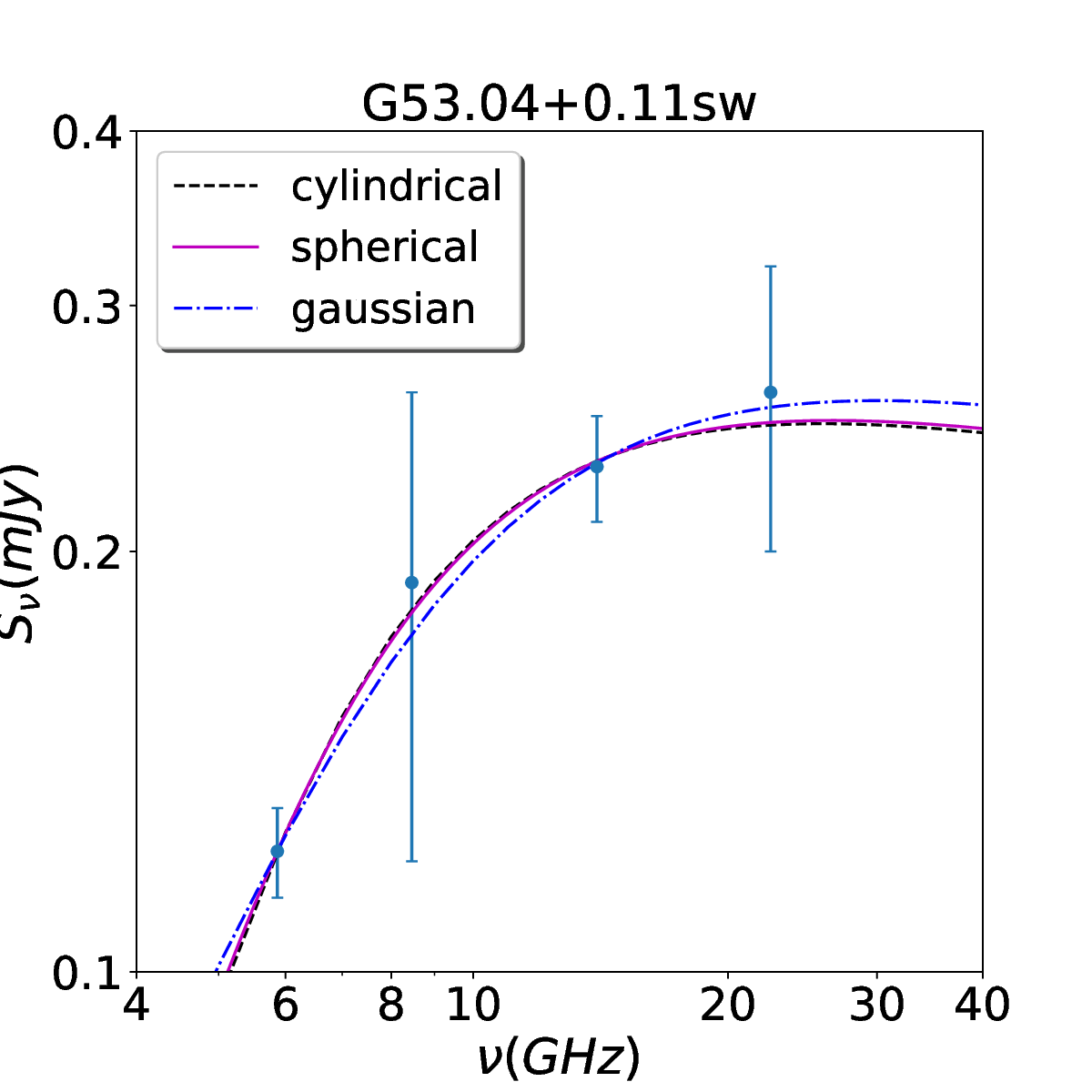}

\vspace{2mm}

\includegraphics[width=0.32\textwidth]{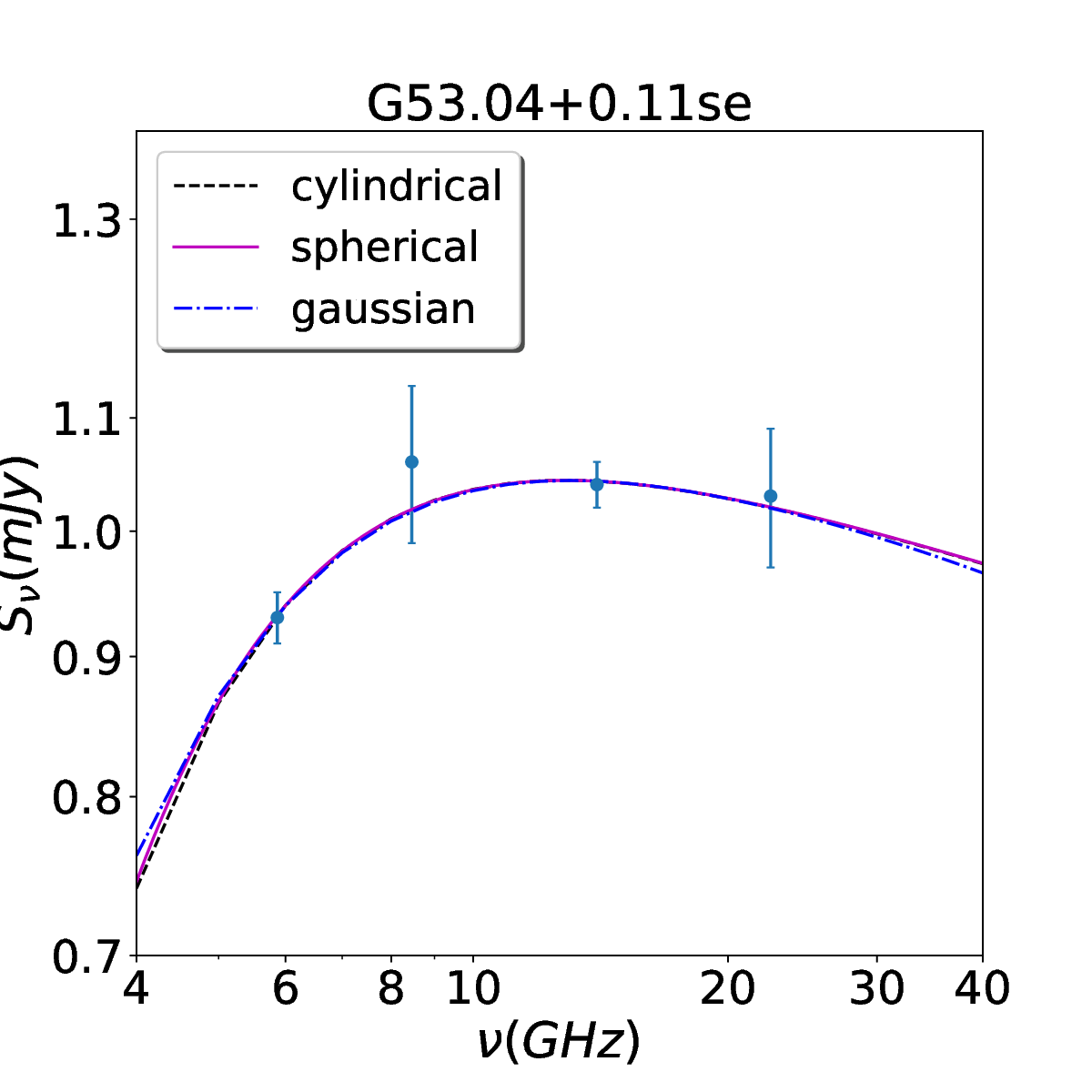}

\caption{Radio continuum spectra. The points correspond to the observations. Models are shown with a dashed line (cylindrical distribution), solid line (spherical distribution), and dot-dashed line (Gaussian distribution).}
\label{esp1}
\end{figure*}

\clearpage

\begin{table*}[hb]
\RaggedRight
\caption{Source Properties and Observing Parameters.}\label{table0}
\begin{center}
\setlength{\tabnotewidth}{\paperwidth}
\setlength{\tabcolsep}{0.6\tabcolsep} \tablecols{15}
\scalebox{0.5}{\begin{tabular}{lcccccccccccccr}
\toprule
Source		&	IRAS		&	\multicolumn{2}{c}{pointing center\tabnotemark{a}}					&	Phase			& 	\multicolumn{2}{c}{6 cm}				&\multicolumn{2}{c}{3.6 cm} 		 	&\multicolumn{2}{c}{2 cm}				&\multicolumn{2}{c}{1.3 cm}					&D\tabnotemark{c}		&	L\tabnotemark{d}		\\
		&	Name		&	$\alpha$(J2000.0)					&	$\delta$(J2000.0)	&	calibrator		&		beam			&	rms		&beam\tabnotemark{b}			&rms		 	&	beam			&	rms		&beam\tabnotemark{b}				&rms				&				&					\\
		&			&								&				&				&		(\arcsec)			&	mJy beam$^{-1}$	&(\arcsec)			&mJy beam$^{-1}$ 	&	(\arcsec)			&	mJy beam$^{-1}$	&(\arcsec)				&mJy beam$^{-1}$		&	kpc			&	10$^{4}$ L$_{\sun}$(kpc)	\\
\midrule
G34.82+0.35	&	 18511+0146	&	18	53	37.90					&	01	50	30.0	&	J1851+0035		&		1.02$\times$0.91	&	0.03		&8.46$\times$7.28	&0.05			&	3.38$\times$1.52	&	0.02		&3.75$\times$2.93		&0.20				&	3.6			&	2.01 (3.86)			\\
G40.28$-$0.22	&	 19031+0621	&	19	05	41.22					&	06	26	12.7	&	J1851+0035		&		0.99$\times$0.89	&	0.03		&8.95$\times$7.34	&0.05			&	3.35$\times$1.47	&	0.02		&2.87$\times$2.23		&0.08				&	4.9			&	2.13 (4.91)			\\
G48.99$-$0.30	&	 19201+1400	&	19	22	26.13					&	14	06	39.8	&	J1922+1530		&		0.94$\times$0.86	&	0.6		&7.18$\times$6.91	&2.0			&	2.93$\times$1.39	&	0.25		&2.66$\times$2.06		&0.30				&	5.1			&	4.5 (5.4)			\\
G53.04+0.11	&	 19266+1745	&	19	28	55.60					&	17	52	03.0	&	J1922+1530		&		0.92$\times$0.85	&	0.02		&7.22$\times$6.69	&0.06			&	2.93$\times$1.37	&	0.02		&3.66$\times$2.44		&0.13				&	9.4			&	5.0 (9.5)			\\
G53.14+0.07	&	 19270+1750	&	19	29	17.58					&	17	56	23.2	&	J1922+1530		&		0.91$\times$0.86	&	0.02		&9.82$\times$7.21	&0.12			&	2.91$\times$1.36	&	0.02		&6.56$\times$6.17		&0.13				&	1.9			&	0.44 (1.8)			\\
\bottomrule
\end{tabular}}
\end{center}
\tabnotetext{a}{Units of right ascension are hours, minutes, and seconds, and units of declination are degrees, arcminutes, and arcseconds.}
\tabnotetext{b}{Beam sizes are from our re-analysis of the \citet{Sanchez2011} data.}
\tabnotetext{c}{Distances from \citet{Pandian2009}, based on methanol masers detected in these regions.}
\tabnotetext{d}{Infrared Luminosity from \citet{Zhang2001}, \citet{Traficante2015}, \citet{Maud2015}, \citet{Lu2014}, \citet{Rathborne2010}. In parentheses we give the distances used to obtain the luminosities.}
\end{table*}

\begin{table*}[hb]
\caption{Observed Parameters of the Continuum Emission.}\label{par-o}
\begin{center}
\setlength{\tabnotewidth}{\paperwidth}
\setlength{\tabcolsep}{0.8\tabcolsep} \tablecols{12}
\scalebox{0.65}{\begin{tabular}{lccccccccccr}
\toprule
Source				&	\multicolumn{2}{c}{Peak Position\tabnotemark{a}}	&	$S_{6cm}$\tabnotemark{b} 	&	$S_{3.6cm}$	&	$S_{2cm}$			&	$S_{1.3cm}$			&	Size\tabnotemark{c}		& R\tabnotemark{d}&R\tabnotemark{d} &	Spectral Index &	T$_b$\tabnotemark{e}\\\cline{2-3}
				&	$\alpha$(J2000.0)	&	 $\delta$(J2000.0)	&	(mJy)				&	(mJy)		&	(mJy)				&	(mJy)				&	\arcsec				&AU                 &mpc&	& K	\\
\midrule
G34.82+0.35w			&	18	53	37.9	&	01	50	30.6	&	0.14$\pm$0.01			&	0.33$\pm$0.02	&	0.26$\pm$0.01			&	$<$1.0				&	$\lesssim$0.51 $\times$0.46 	&900			 &4&	0.6$\pm$0.4	&$>$24	\\
G34.82+0.35e			&	18	53	38.68	&	01	50	13.6	&	0.42$\pm$0.02			&	0.66$\pm$0.05	&	0.64$\pm$0.01			&	$<$1.0				&	$\lesssim$0.51 $\times$0.46 	&900               	 &4&	0.4$\pm$0.2	&$>$73	\\
G40.28$-$0.22			&	19	05	41.21	&	06	26	13.0	&	0.15$\pm$0.08			&	0.75$\pm$0.07	&	0.63$\pm$0.04			&	1.4$\pm$0.2			&	$\lesssim$0.49$\times$0.45	&1200              	 &6&	1.64$\pm$0.07	&$>$42	\\
G48.99$-$0.30			&	19	22	26.13	&	14	06	39.7	&	5.68$\pm$0.12			&	7.7$\pm$0.6	&	9.0$\pm$0.2			&	9.4$\pm$0.7			&	$\lesssim$0.47$\times$0.43	&1100		    	 &6&	0.48$\pm$0.07	&$>$1100	\\
G53.04+0.11nw			&	19	28	55.14	&	17	52	09.7	&	0.36$\pm$0.03			&	0.46$\pm$0.02	&	0.52$\pm$0.02			&	$<$0.36				&	$\lesssim$0.46$\times$0.43	&2100		    	 &10&	0.42$\pm$0.05	&$>$75	\\
G53.04+0.11sw			&	19	28	55.49	&	17	52	03.1	&	0.12$\pm$0.01			&	0.19$\pm$0.07	&	0.23$\pm$0.02			&	0.26$\pm$0.06			&	$\lesssim$0.46$\times$0.43	&2100			 &10&	0.71$\pm$0.07	&$>$25	\\
G53.04+0.11se			&	19	28	55.68	&	17	52	03.3	&	0.93$\pm$0.02			&	1.06$\pm$0.07	&	1.04$\pm$0.02			&	1.03$\pm$0.06			&	$\lesssim$0.46$\times$0.43	&2100			 &10&	0.12$\pm$0.03	&$>$190	\\
G53.14+0.07sw\tabnotemark{f}	&	19	29	17.55	&	17	56	18.8	&	$<$0.1				&	$<$0.60		&	0.30$\pm$0.05			&	3.3$\pm$0.4\tabnotemark{g}	&	3.59$\times$1.10		&1900			 &9&	-		&0.5	\\
G53.14+0.07n\tabnotemark{f}	&	19	29	17.58	&	17	56	23.4	&	$<$0.1				&	$<$0.60		&	0.16$\pm$0.01			&	-				&	$\lesssim$1.45$\times$0.68	&940			 &5&	-			&$>$1.1	\\
G53.14+0.07se\tabnotemark{f}	&	19	29	17.82	&	17	56	20.2	&	$<$0.1				&	$<$0.60		&	0.17$\pm$0.01			&	-				&	$\lesssim$1.45$\times$0.68	&940			 &5&	-			&$>$1.2	\\
\bottomrule
\end{tabular}}
\end{center}
\tabnotetext{a}{Units of right ascension are hours, minutes, and seconds, and units of declination are degrees, arcminutes, and arcseconds. Positions are obtained from the 6 cm images except for the G53.14+0.07 region for which the 2 cm image was used.}
\tabnotetext{b}{The flux density limit for non-detections is set to $3\sigma$, where $\sigma$ is the rms noise of the maps given in Table \ref{table0}.}
\tabnotetext{c}{We use 6 cm maps to estimate the source sizes. If the sources are unresolved we use half the synthesized beam size. If the sources are resolved we report the size deconvolved from the beam.}
\tabnotetext{d}{The radius is estimated as half of the geometric mean of the major and minor axes given in the Size column and using the source distance from Table \ref{table0}.}
\tabnotetext{e}{Brightness temperatures were calculated using the reported source size and 6 cm flux density, except for G53.14+0.07 for which the 2 cm values were used.}
\tabnotetext{f}{For this source we use the 2 cm image for the position, size and brightness temperature.}
\tabnotetext{g}{Only one source is detected at 1.3 cm owing to the lower angular resolution. The 1.3 cm flux shown for this source is the flux of the more extended region.}

\end{table*}

\begin{table*}[hb]
\caption{Calculated parameters using Spherical Olnon Model.}\label{par-aj}
\begin{center}
\setlength{\tabnotewidth}{\linewidth}
\setlength{\tabcolsep}{0.9\tabcolsep} \tablecols{12}
\scalebox{0.7}{\begin{tabular}{lccccccccccc}
\toprule
Source		&	       R	        &	R	& 	  R	&	     Ne	 		&  EM			& \ \ $\tau$ \tabnotemark{a}	&	Mi			&	Ni			&	logNi		&SP\tabnotemark{b}	&L\tabnotemark{b}		\\
		&	       mpc	        &	AU	&       \arcsec	&      10$^{5}$ cm$^{-3}$	& 10$^{8}$ pc cm$^{-6}$	&				&	10$^{-6}$M$_{\sun}$	&	10$^{46}$ phot $s^{-1}$	&			&			&10$^{4}$ L$_{\sun}$		\\
\midrule
G34.82+0.35w	&	   0.3$\pm$  0.1	&	62	&	0.02	&	   8 $\pm$4 		&	3.8		&	2.6,1.2,0.4 ,-- 	&	1.7			&	0.055			&	44.74		&	B2		&		0.29		\\
G34.82+0.35e	&	   0.50$\pm$ 0.05	&	100	&	0.03	&	   5 $\pm$1 		&	2.5		&	1.8,0.8,0.3,--   	&	6.2			&	0.10			&	45.00		&	B1		&		0.50		\\
G40.28$-$0.22	&	   0.32$\pm$ 0.05	&	66	&	0.02	&	   24$\pm$10 		&	37		&	29.1,13.5,4.7,1.7	&	7.8			&	0.60			&	45.78		&	B0.5		&	        1.10  		\\
G48.99$-$0.30	&	   2.48$\pm$ 0.01 	&	510	&	0.10	&	   2.32 $\pm$ 0.02 	&	2.7		&	2.1, 0.9, 0.3, 0.1  	&	369.0			&	2.63			&	46.42		&	B0.5		&		1.10		\\
G53.04+0.11nw	&	   1.22$\pm$ 0.01	&	251	&	0.03	&	   2.94$\pm$ 0.02 	&	2.1		&	1.7,  0.8,0.3,--   	&	55.7			&	0.50			&	45.70		&	B0.5		&		1.10		\\
G53.04+0.11sw	&	   0.63$\pm$0.01	&	130	&	0.01	&	   5.5$\pm$0.1		&	3.8		&	3.0,1.4,0.5,0.2		&	14.2			&	0.24			&	45.38		&	B1		&		0.50		\\
G53.04+0.11se	&	   2.6 $\pm$0.1		&	536	&	0.058	&	   1.3$\pm$ 0.07 	&	0.88		&	0.7,0.3,0.1,0.04 	&	244.5			&	0.95			&	45.98		&	B0.5		&		1.10		\\
\bottomrule
\end{tabular}}
\end{center}
\tabnotetext{a}{$\tau$ calculated from the spherical model for each wavelength (C, X, U, K) with detection (see Table \ref{par-o}).}
\tabnotetext{b}{Spectral type (SP) and luminosity (L) based on $\log N_i$ from \citet{Panagia1973}.}
\end{table*}

\begin{table}[hb]
\caption {Mass loss rate for the spherical stellar wind model.}\label{wind}
\begin{center}
\setlength{\tabnotewidth}{\linewidth}
\setlength{\tabcolsep}{1.2\tabcolsep} \tablecols{7}
\scalebox{0.9}{\begin{tabular}{lcccccc}
\toprule
Sources	&	$\dot{M}$\tabnotemark{a}	&	N$_i$\tabnotemark{b}	&	Radius\tabnotemark{b}	&	$\dot{M_{*}}$\tabnotemark{c}	&	L\tabnotemark{b}	&L\tabnotemark{d} 		\\
	&	10$^{-5}$M$_{\sun}$ yr$^{-1}$		&	fot.s$^{-1}$	&	R$_{\sun}$		&	10$^{-5}$M$_{\sun}$ yr$^{-1}$	&	10$^{4}$ L$_{\sun}$	&10$^{4}$ L$_{\sun}$			\\
\midrule
G34.82+0.35w	&	0.59			&	2.3$\times$10$^{47}$	&	5.5			&	0.66				&	2.5			&2.0		\\
G34.82+0.35e	&	1.33			&	6.9$\times$10$^{47}$	&	6.0			&	1.20				&	3.8			&2.0		\\
G40.28$-$0.22	&	0.98			&	6.9$\times$10$^{47}$	&	6.0			&	1.20				&	3.8			&2.13		\\
G48.99$-$0.30	&	15.9			&	4.2$\times$10$^{49}$	&	12.6			&	13.5				&	67.6			&4.5		\\
G53.04+0.11nw	&	5.02			&	6.6$\times$10$^{48}$	&	8.1			&	4.3				&	14.8			&5		\\
G53.04+0.11sw	&	2.22			&	2.2$\times$10$^{47}$	&	6.5			&	2.2				&	6.5			&5		\\
G53.04+0.11se	&	10.2			&	2.3$\times$10$^{49}$	&	10.7			&       9.2  	   		        &	39.8			&5		\\
G53.14+0.07sw	&	0.26			&	1.7$\times$10$^{46}$	&	5.1			&	0.17				&	1.1			&0.44		\\
G53.14+0.07n	&	0.16			&	1.7$\times$10$^{46}$	&	5.1			&	0.17				&	1.1 			&0.44		\\
G53.14+0.07se	&	0.17			&	1.7$\times$10$^{46}$	&	5.1			&	0.17 				&	1.1 			&0.44		\\
\bottomrule
\end{tabular}}
\end{center}
\tabnotetext{a}{Mass loss rate ($\dot{M}$) determined from equation \ref{flux}, using the flux density at 5.868~GHz (Table \ref{par-o}); distances are from Table \ref{table0}; the terminal velocity is 1000~km~s$^{-1}$. For G53.14+0.07 the 2~cm (14.5~GHz) flux density was used.}
\tabnotetext{b}{Ionizing photon rate ($N_i$), radius ($R_{\sun}$) and luminosity (L) from \citet{Panagia1973}.}
\tabnotetext{c}{Maximum mass loss rate ($\dot{M}_{*}$) determined from equation \ref{ion}, using $N_i$ and $R$ from \citet{Panagia1973} and a terminal velocity of 1000~km~s$^{-1}$.}
\tabnotetext{d}{Infrared luminosity (L) from Table \ref{table0}.}
\end{table}

\begin{table}[hb]
\caption{Mass loss rate for Jet}\label{jet}
\begin{center}
\setlength{\tabnotewidth}{10cm}
\setlength{\tabcolsep}{0.8\tabcolsep}\tablecols{3}
\scalebox{0.9}{\begin{tabular}{lccr}
\toprule
Sources		&	$\dot{M_{c}}$\tabnotemark{a}		&	$\dot{M_{c}}$\tabnotemark{a}	       &$\alpha_{op}$\tabnotemark{b}	\\
		&	10$^{-6}$ M$_{\odot}$ yr$^{-1}$		&	10$^{-6}$ M$_{\odot}$ yr$^{-1}$	       &				\\
		&$i$=45\textdegree and $\theta$=27\textdegree	&$i$=25\textdegree and $\theta$=10\textdegree  &				\\
\midrule
G34.82+0.35w	&	9					&	5				&0.6$\pm$0.4		\\
G34.82+0.35e	&	14					&	7				&0.4$\pm$0.3		\\
G40.28$-$0.22	&	--					&	--				&1.6$\pm$0.1		\\
G48.99$-$0.30	&	393					&	202				&0.8$\pm$0.1		\\
G53.04+0.11nw	&	89					&	46				&0.7$\pm$0.1		\\
G53.04+0.11sw	&	295					&	151				&1.2$\pm$0.2		\\
G53.04+0.11se	&	94					&	48				&0.4$\pm$0.1		\\
\bottomrule
\end{tabular}}
\end{center}
\tabnotetext{a}{Mass loss rate obtained from equation \ref{csw}.}
\tabnotetext{b}{$\alpha_{op}$ is the opaque spectral index.}
\end{table}

\begin{table}[hb]
\caption{Parameters of clumps associated with the regions\tabnotemark{a}.}\label{table6}
\begin{center}
\setlength{\tabnotewidth}{8cm}
\setlength{\tabcolsep}{0.8\tabcolsep} \tablecols{4}
\scalebox{0.9}{\begin{tabular}{lccr}
\toprule
Region		&	T$_{kin}$(NH3)	&	  F$_{int}$(870~$\mu$m)	&  Clump Mass  		\\
		&	K		&		Jy		&  M$_{\odot}$ \\
\midrule
G34.82+0.35	&	18.9		&	 13.21			&	1000		\\
G40.28$-$0.22	&	30.8		&	 13.24			&	4200		\\
G48.99$-$0.30	&	27.6		&	 76.70			&	12000		\\
G53.04+0.11	&	25.5		&	  4.95			&	2800		\\
G53.14+0.07	&	22.6		&	 15.35			&	220		\\
\bottomrule
\end{tabular}}
\end{center}
\tabnotetext{a}{All values are from \citet{Urquhart2011, Urquhart2014} except for the kinematic temperature of G40.28$-$0.22, which is from \citet{Cyganowski2013}. The clump masses assume a dust temperature of 20~K.}
\end{table}

\begin{table*}[ht!]
\caption{Tracers of star formation in the sample.}\label{tracers}
\begin{center}
\setlength{\tabnotewidth}{\linewidth}
\setlength{\tabcolsep}{11pt}\tablecols{11}
\scalebox{0.8}{\begin{tabular}{lcccccccccr}
\toprule
Source			&	EGO\tabnotemark{a}	&H$_{2}$O\tabnotemark{b}	&CH$_{3}$OH\tabnotemark{c}	&CH$_{3}$OH\tabnotemark{d}	&	Spitzer\tabnotemark{e}	&	1.1 mm\tabnotemark{f}		& \multicolumn{4}{c}{Radio Continuum}				\\
				&		&		&Class II	&Class I	&		&		&			C\tabnotemark{h}	&	X\tabnotemark{g}	& KU\tabnotemark{h}		& K\tabnotemark{g}		\\
\midrule
G34.82+0.35w			&	-	&	y	&	y	&	y	&	y	&	y	&		y	&	y	&	y	&	n	\\
G34.82+0.35e			&	-	&	y	&	y	&	n	&	y	&	y	&		y	&	y	&	y	&	y	\\
G40.28$-$0.22			&	y	&	y	&	y	&	y	&	y	&	y	&		y	&	y	&	y	&	y	\\
G48.99$-$0.30			&	-	&	y	&	y	&	-	&	y	&	y	&		y	&	y	&	y	&	y	\\
G53.04+0.11			&	-	&	y	&	y	&	y	&	y	&	y	&		y	&	y	&	y	&	y	\\
G53.14+0.07			&	-	&	y	&	y	&	y	&	y	&	y	&		n	&	n	&	y	&	y	\\
\bottomrule
\end{tabular}}
\end{center}
\tabnotetext{a}{Extended Green Object (EGO) \citet{Cyganowski2008}.}
\tabnotetext{b}{H$_2$O masers \citet{Urquhart2011, Cyganowski2013, Nagayama2015, Sridharan2002}.}
\tabnotetext{c}{CH$_3$OH class~II masers \citet{Pandian2011, Szymczak2012, Sun2014}.}
\tabnotetext{d}{CH$_3$OH class~I maser \citet{Kurtz2004, Chen2011, Litovchenko2011, Chen2012}.}
\tabnotetext{e}{Infrared emission \citet{Vig2007, Cyganowski2008, Sun2014} and GLIMPSE/MIPSGAL.}
\tabnotetext{f}{1.1~mm emission \citet{Schlingman2011, Rosolowsky2010, Beuther2002a}.}
\tabnotetext{g}{1.3 and 3.6~cm emission \citet{Sanchez2011}.}
\tabnotetext{h}{2 and 6~cm emission (this work).}

\end{table*}

\clearpage

\begin{appendices}
\section{Appendix}
\label{appe}
\subsection*{HII regions}

We applied the models developed by \citet{Olnon1975}, who assumed ionized hydrogen gas, circular symmetry around  the line of sight and  uniform electron temperature ($T_{e}$). In the Rayleigh-Jeans regime, the total flux can be expressed by

\begin{equation}
S_{\nu}=\frac{4 \, \pi k \, T_{e} \, \nu^{2}}{c^{2} \, D^{2}}\int_{0}^{\infty}\rho \left[1-e^{-\tau_{\nu}(\rho)}\right] d \rho
\end{equation}

To calculate the optical depth of the region, we need to know its geometry and distance. One way to do this is to use the parameter $\rho$, which is the radial distance from the center of the region to a point on its surface, measured in a plane perpendicular to the line of sight. The distance D is the distance from us to the region. The optical depth at a frequency $\nu$ is given by $\tau_{\nu}(\rho)= 0.08235 \times \, T_{e}^{-1.35} \, \nu^{-2.1} E(\rho)$, where $T_e$ is the electron temperature and $E(\rho)$ is the emission measure. The emission measure depends on the electron density profile of the region, and can be written as $E(\rho)=2 \int_{0}^{\infty}n_{e}^{2}(r)dz$ , where $r^{2}=\rho^{2}+z^{2}$ and $z$ is the coordinate along the line of sight. We tested different models for the electron density distribution, such as cylindrical, Gaussian, and spherical, following the method of \citet{Olnon1975}.

In what follows we adopt the notation introduced by \citet{Olnon1975}. Specifically, we write $p$ instead of the central optical depth, $\tau_\nu(0) = fE(0)$.
Here $f(\nu, T_e)$ is the Altenhoff approximation \citep{Olnon1975} for the free-free absorption coefficient, given by
\begin{equation}
f=f(\nu, T_e) = 8.235 \times 10^{-2} 
\left( \frac{T_e}{\mathrm{K}} \right)^{-1.35} 
\left( \frac{\nu}{\mathrm{GHz}} \right)^{-2.1} 
\, \mathrm{cm}^6 \, \mathrm{pc}^{-1}.
\end{equation}
These definitions and approximations are included here to clarify the notation used in the subsequent equations.

\textbf{The cylindrical distribution.}

Following \citet{Olnon1975}, we assume a cylindrical geometry in which the electron density is constant, $n_e = n_0$, inside a cylinder of radius $R$ and length $2R$, with its axis of symmetry aligned along the line of sight. Outside the cylinder, $n_e = 0$.

\begin{equation}
p = 2 n_0^2 R f
\end{equation}

\begin{equation}
 S_{\nu}=\frac{2 \, \pi k \, T_{e} R^{2} \, \nu ^{2}}{C^{2} \, D^{2}} \left(1-e^{-p}\right)
\end{equation}

\textbf{The Gaussian distribution }

Following \citet{Olnon1975}, we adopt spherical symmetry and assume a Gaussian distribution for the electron density, given by
\begin{equation}
n_e(r) = n_0 \exp\left(-\frac{r^2}{2R^2}\right),
\end{equation}
where $n_0$ is the central electron density and $R$ characterizes the scale of the distribution.

\begin{equation}
p = n_0^2 R f \sqrt{\pi}
\end{equation}

\begin{equation}
 S_{\nu}=\frac{2 \, \pi k \, T_{e} R^{2} \, \nu ^{2}}{C^{2} \, D^{2}} \left[\gamma+ln \, p+E_{1}(p)\right]
\end{equation}

Where  The function $E_1(p)$ denotes the exponential integral evaluated between $p$ and infinity,
\begin{equation}
E_1(p) = \int_{p}^{\infty} \frac{e^{-t}}{t}\, dt,
\end{equation}
and $\gamma$ is the Euler–Mascheroni constant ($\gamma \approx 0.5772$).

\textbf{The spherical distribution}

Following \citet{Olnon1975}, we  consider a spherical geometry in which the electron density is uniform, $n_e = n_0$, inside a sphere of radius $R$, and $n_e = 0$ outside the sphere.

\begin{equation}
p = 2 n_0^2 R f
\end{equation}

\begin{equation}
 S_{\nu}=\frac{2 \, \pi k \, T_{e} R^{2} \, \nu ^{2}}{C^{2} \, D^{2}} \left[1-\frac{2}{p^{2}}{1-(p-1)e^{-p}}\right]
\end{equation}

We applied the minimizing function in Python software to find the optimal values for radius and density by fitting the different models to the data using the least-squares method. The equation we used for the fit is $\chi^{2}=\sum_{i}^{N}\frac{\left[S_{\nu i}^{obs}-S_{\nu
i}^{mod}(a)\right]^{2}}{ \epsilon_{i}^{2}}$, where $S_{\nu i}^{obs}$  represents the observed data, $S_{\nu i}^{mod}$ is the model, $a$ is the set of parameters that we want to optimize, and $\epsilon_{i}$ is the
error in the flux density (see table \ref{par-o}).

\end{appendices}

\bibliographystyle{plainnat}
\bibliography{biblio}

\end{document}